\documentclass[12pt]{article}
\usepackage{amsmath, amssymb, graphicx, hyperref}
\usepackage{geometry}
\usepackage{pdflscape}
\usepackage{booktabs}
\usepackage{siunitx}
\usepackage{caption}
\usepackage{graphicx}
\usepackage{subcaption}
\usepackage{appendix}
\usepackage{mathtools}
\usepackage{array}
\newcolumntype{C}{>{\centering\arraybackslash}p{1.2cm}}

\title{\textbf{The Hidden Constant of Market Rhythms: How $1 - 1/e$ Defines Scaling in Intrinsic Time}}
\author{Thomas Houweling}
\date{\today}

\begin{document}
\maketitle

\begin{abstract}
Directional-change Intrinsic Time analysis has long revealed scaling laws in market microstructure, but the origin of their stability remains elusive. This article presents evidence that Intrinsic Time can be modeled as a memoryless exponential hazard process. Empirically, the proportion of directional changes to total events stabilizes near $1 - e^{-1} \approx 0.632$, matching the probability that a Poisson process completes one mean interval. This constant provides a natural heuristic to identify scaling regimes across thresholds and supports an interpretation of market activity as a renewal process in intrinsic time.
\end{abstract}

\section{Introduction}
\label{sec:intro}
The \emph{Intrinsic Time} paradigm defines time through an event-based and algorithmic mechanism, providing a foundation for a wide range of applications in economic modeling, forecasting, and trading \cite{aloud2011directional, glattfelder2025modern, ma2017volatility, petrov2019instantaneous, petrov2019multidimensional, petrov2020agent}. A recent overview of its theoretical development and applications is provided in \cite{glattfelder2024theory}. At the heart of the framework lies the \emph{Directional-change/Overshoot} (DcOS) operator, which classifies price movements into two distinct event types. A \emph{Directional-change} (Dc) occurs when the series reverses by a specified threshold $\delta$ from a local extremum, whereas an \emph{Overshoot} (OS) occurs when it continues in the same direction by another $\delta$. 

Because event-defining thresholds are, in principle, unbounded, applying multiple DcOS operators with varying $\delta$ values to the same time series enables multi-scale analysis—from microscopic to macroscopic temporal resolutions. Empirical studies have shown that financial price dynamics exhibit multiple scaling laws, characteristic of self-organizing complex systems \cite{guillaume1997birds, glattfelder2011patterns}. In particular, event frequencies follow power-law relationships with respect to the DcOS threshold $\delta$. However, the mechanisms underlying this self-similar structure of market time remain poorly understood. One objective of this work is to investigate the processes responsible for the observed scaling behavior.

A second objective is to develop a practical tool for identifying an appropriate range of $\delta$ thresholds in multi-scale DcOS analyses. Because $\delta$ is theoretically unbounded, selecting meaningful threshold values has long been a methodological challenge for Intrinsic Time practitioners. This work proposes a heuristic approach to constrain and interpret the empirically relevant range of $\delta$ thresholds.

The remainder of this article is organized as follows. \S\ref{subsec:renewal} and \S\ref{subsec:sixty_three} develop the renewal-process theory and its link to the $1-e^{-1}$ constant. \S\ref{sec:methods} describes our statistical diagnostics. \S\ref{sec:results} presents empirical results from three cryptocurrency datasets. \S\ref{sec:discussion} interprets the findings and discusses practical applications. Conclusions are in \S\ref{sec:conclusion}.

\subsection{Intrinsic Time as a Renewal Process}
\label{subsec:renewal}

\paragraph{Motivation.}
Under the \emph{Directional-change/Overshoot} (DcOS) decomposition, each time a \emph{Directional-change} (Dc) event occurs, the reference extremum is reset and a new event cycle begins. 
This reset property implies that the evolution of the price path \emph{after} a Dc event depends only on the new extremum and not on the overshoot history that preceded it. 
In this sense, the sequence of Dc events provides a natural set of ``renewal epochs'' at which the process restarts.  
This motivates modelling intrinsic time as a \emph{renewal process}.

\paragraph{Renewal-process formalism.}
Let
\[
0 = \tau_0 < \tau_1 < \tau_2 < \cdots
\]
denote the successive times at which a Directional-change occurs.  
These times are the \emph{renewal epochs}.  
Define the inter-renewal durations as
\begin{equation}
T_n = \tau_n - \tau_{n-1}, \qquad n \ge 1 .
\end{equation}
A renewal process assumes that the random variables $(T_n)_{n\ge1}$ are i.i.d.\ nonnegative with cumulative distribution function
\begin{equation}
F(t) = \Pr(T \le t) ,
\end{equation}
and density $f(t) = F'(t)$ when it exists.  
The associated counting process
\begin{equation}
N(t) = \max\{ n \ge 0 : \tau_n \le t \}
\end{equation}
counts the number of renewals (i.e., the number of Directional-changes) up to intrinsic time $t$.  

The inter-renewal distribution is fully described by its \emph{hazard rate}
\begin{equation}
h(t) = \frac{f(t)}{1 - F(t)} ,
\end{equation}
which measures the instantaneous probability of a renewal occurring at time $t$, conditional on no renewal having occurred before $t$.

\paragraph{Memoryless specialization.}
If the DcOS mechanism is approximately memoryless after each reset, then the inter-renewal duration $T$ is exponential,
\begin{equation}
T \sim \mathrm{Exp}(\lambda),
\end{equation}
with distribution function
\begin{equation}
\Pr(T \le t) = 1 - e^{-\lambda t},
\end{equation}
and the defining memoryless property
\begin{equation}
\Pr(T > s + t \mid T > s) = \Pr(T > t).
\end{equation}
In words, the probability of waiting at least another \emph{t} unit of time for a Dc event to occur conditional on having already waited \emph{s} without observing a Dc event, is equal to the unconditional probability of waiting more than \emph{t} from the very start. In short, past waiting does not change the distribution of the additional waiting time. In this case the hazard rate is constant,
\begin{equation}
h(t) \equiv \lambda ,
\end{equation}
and the renewal counting process $N(t)$ becomes a Poisson process with rate~$\lambda$.  
The exponential law is the \emph{only} continuous distribution with the memoryless property, which makes it the canonical choice for models in which the system resets fully after each event.

\paragraph{Canonical examples.}
Exponential renewals arise throughout the sciences: exponential lifetimes in reliability theory~\cite{Balakrishnan2022Robust}, Poisson arrivals in queueing systems~\cite{Kleinrock1975Queueing}, radioactive decay in nuclear physics~\cite{Peshkin2017NonExponential}, and first-order dynamical responses in engineering~\cite{Elmore1948Transient}.  
In all these settings, past history is irrelevant once the system has reset, mirroring the logic of DcOS cycles.

\paragraph{Why renewal structure matters for intrinsic time.}
Interpreting DcOS cycles as renewals provides several benefits:
\begin{enumerate}
    \item It yields \emph{testable distributional predictions}.  
          In particular, memorylessness implies exponential overshoot-length distributions and geometric overshoot-count distributions.
    \item It links \emph{frequency scaling} to threshold choice.  
          If total event frequency scales as $\delta^{-\beta}$ across a range of thresholds, then Directional-change frequencies inherit the same exponent~$\beta$.
    \item It produces \emph{constant success-rate predictions}.  
          A constant hazard rate implies that the proportion of Directional-change events among all DcOS events is constant across thresholds within the scaling regime.
\end{enumerate}
These implications provide direct hypotheses that can be tested empirically, which motivates the next subsection.

\subsection{The Meaning of the \texorpdfstring{$1 - e^{-1}$}{1 - e\string^-1} Constant}
\label{subsec:sixty_three}

In the exponential renewal model, the probability that a renewal occurs within one mean waiting time is
\begin{equation}
\Pr(T \le 1/\lambda) = 1 - e^{-1},
\end{equation}
a universal constant equal to approximately $0.6321$.  
This quantity appears in all first-order, memoryless systems \cite{Elmore1948Transient, Kleinrock1975Queueing}: it is the fraction of the stationary response achieved after one time constant in engineering, the survival complement after one mean lifetime in reliability theory, and the success probability after one expected waiting period in exponential stopping-time problems.

\paragraph{Mapping the constant to DcOS events.}
A DcOS cycle begins with a Directional-change, continues through a number of Overshoots, and terminates at the next Directional-change.  
Let $nDc$ denote the total number of Directional-change events and $n_{\mathrm{OS}}$ the total number of Overshoot events across many cycles.  
If overshoot lengths are exponential in units of the threshold~$\delta$, then the number of Overshoots per cycle follows a geometric distribution.

Let $X$ be the (normalized) overshoot length, and assume
\begin{equation}
X \sim \mathrm{Exp}(\lambda).
\end{equation}
Discretizing $X$ into integer multiples of $\delta$ produces an overshoot-count random variable
\begin{equation}
K \sim \mathrm{Geom}(p),
\end{equation}
with parameter
\begin{equation}
p = 1 - e^{-\lambda},
\end{equation}
and mean
\begin{equation}
\mathbb{E}[K] = \frac{1-p}{p}.
\end{equation}

Over many cycles, the observed fraction of Directional-change events among all DcOS events satisfies
\begin{equation}
\frac{n_{\mathrm{Dc}}}{n_{\mathrm{Dc}} + n_{\mathrm{OS}}}
\;\xrightarrow[]{\;\text{LLN}\;}
\frac{1}{1 + \mathbb{E}[K]}
= p
= 1 - e^{-\lambda}.
\end{equation}
Thus, if DcOS cycles behave like exponential renewals with $\lambda \approx 1$, then the Directional-change share should be close to
\begin{equation}
\frac{n_{\mathrm{Dc}}}{n_{\mathrm{Dc}}+n_{\mathrm{OS}}} \approx 1 - e^{-1} \approx 0.6321.
\end{equation}

\paragraph{Testable predictions from the renewal hypothesis.}
If DcOS truly behaves as a renewal process over a given threshold range, then:
\begin{enumerate}
    \item The ratio $n_{\mathrm{Dc}} / (n_{\mathrm{Dc}} + n_{\mathrm{OS}})$ should be approximately constant across $\delta$ and close to $1 - e^{-\lambda}$.
    \item The overshoot-count distribution $K$ should be geometric with parameter
    
    $p \approx n_{\mathrm{Dc}}/(n_{\mathrm{Dc}}+n_{\mathrm{OS}})$.
    \item The normalized overshoot-length distribution should be exponential with rate~$\lambda$.
    \item Dc event frequencies should scale as
          \begin{equation}
          \mathrm{DcFreq}(\delta) \approx p \cdot C \, \delta^{-\beta},
          \end{equation}
          inheriting the same exponent $\beta$ as total event frequencies.
\end{enumerate}

\paragraph{Relevance and caveats.}
The benchmark value $1 - e^{-1}$ provides a practical diagnotic for identifying the threshold region in which DcOS behaves like a memoryless process.  
Values significantly below this constant typically reflect microstructure effects or overshoot persistence, while values above it suggest undersampling at large thresholds.  
In empirical applications, small deviations (e.g., Directional-change shares around $0.60$–$0.62$) arise naturally from mild dependence or volatility regime-mixing.  
These deviations can be interpreted as a renormalized hazard rate $\lambda_{\mathrm{eff}} < 1$ via the identity
\begin{equation}
p = 1 - e^{-\lambda_{\mathrm{eff}}}.
\end{equation}

The next section describes how these predictions are tested empirically using Directional-change counts, Overshoot counts, and overshoot-length distributions.

\section{Methods}
\label{sec:methods}
This section describes the statistical diagnostics used to assess whether the \emph{Directional-change/Overshoot} (DcOS) decomposition of a price series exhibits memoryless (renewal-type) behavior and scale-consistent (power-law) event frequencies. 
All tests are applied to counts and event logs computed over a grid of Directional-change thresholds $\delta$ spanning multiple orders of magnitude. 
Analyses were run with the \texttt{IntrinsicTime} Python package (v1.3; \href{https://pypi.org/project/IntrinsicTime/}{PyPI}, \href{https://github.com/THouwe/IntrinsicTime}{GitHub}).%
\footnote{For reproducibility, we refer to the publicly available source at \href{https://github.com/THouwe/IntrinsicTime}{github.com/THouwe/IntrinsicTime} and the packaged release at \href{https://pypi.org/project/IntrinsicTime/}{pypi.org/project/IntrinsicTime/}.}

We evaluate three one‑second sampled midprice cryptocurrency datasets. 
Table~\ref{tab:datasets} summarizes dataset identifiers, date ranges, and calendar-day coverage.\footnote{Calendar days are counted inclusively, i.e., both start and end dates are included.}

\begin{table}[h!]
\centering
\caption{Datasets used in the study: identifiers, date ranges, and calendar-day coverage.}
\label{tab:datasets}
\begin{tabular}{lll r}
Dataset & Instrument & Date range (YYYY--MM--DD) & Days \\
\midrule
A & BTCUSDT & 2022-02-08 to 2023-03-24 & 410 \\
B & BTCUSDT & 2023-03-24 to 2024-07-01 & 466 \\
C & ETHUSDT & 2023-12-27 to 2024-08-09 & 227 \\
\end{tabular}
\end{table}

For each dataset, we compute the Intrinsic Time representation of the midprice series using a grid of fifty logarithmically spaced Directional-change thresholds $\delta \in [10^{-5},\,1]$. This wide range, from values close to or below instrument precision up to a 100\% price move, is intentional, allowing us to probe the limits where the method remains valid. Directional-change and Overshoot events are extracted using the \texttt{dcos\_core} module, while event counts across thresholds are generated through the \texttt{dcos\_fractal} module. 
Statistical diagnOStics for the renewal-process hypothesis are implemented in the \texttt{dcos\_tests} module, which evaluates distributional properties of overshoot counts, overshoot lengths, and Directional-change shares.

For notation, let $n_{\mathrm{Dc}}(\delta)$, $n_{\mathrm{OS}}(\delta)$, and $n_{\mathrm{Ev}}(\delta)$ denote, respectively, the total number of Directional-change events, Overshoot events, and all DcOS events at threshold $\delta$, with
\begin{equation}
n_{\mathrm{Ev}}(\delta) \;=\; n_{\mathrm{Dc}}(\delta) + n_{\mathrm{OS}}(\delta).
\end{equation}
Unless stated otherwise, event \emph{frequencies} refer to counts normalized by the number of ticks in the sample. 
Subsequent subsections detail the renewal-based hypotheses and the associated tests (Directional-change share stability, geometric overshoot-count fit, exponential overshoot-length fit, and shared scaling exponents across $\delta$).

\subsection{Empirical Directional-change Probability}
\label{sec:p_mean}

For each threshold $\delta$, the total number of directional changes ($n_{\mathrm{Dc}}$)
and overshoots ($n_{\mathrm{OS}}$) was recorded.  
The empirical directional-change probability was computed as:
\begin{equation}
p_1 = \frac{n_{\mathrm{Dc}}}{n_{\mathrm{Dc}} + n_{\mathrm{OS}}}, \qquad
p_2 = \frac{1}{1 + \frac{n_{\mathrm{OS}}}{n_{\mathrm{Dc}}}}.
\end{equation}
The two definitions are algebraically equivalent under ideal counting, and their mean
\begin{equation}
\bar{p} = \frac{1}{2}\left(p_1 + p_2\right)
\end{equation}
was used as a robust estimator of the intrinsic directional-change probability.

The standard error of $\bar{p}$, assuming binomial sampling, was estimated as:
\begin{equation}
\mathrm{SE}(\bar{p}) = \sqrt{ \frac{\bar{p}(1-\bar{p})}{n_{\mathrm{Dc}} + n_{\mathrm{OS}}} }.
\end{equation}

This empirical probability was compared against the theoretical constant
\begin{equation}
p_0 = 1 - e^{-1} \approx 0.6321,
\end{equation}
which arises from a memoryless exponential hazard model
(\S\ref{sec:exp_test}).  The deviation
\[
\Delta p = \bar{p} - p_0
\]
serves as a scale-dependent indicator of the degree to which the DcOS event process
conforms to renewal dynamics (Table \ref{tab:empirical_p}).

\begin{table}[h!]
\centering
\caption{Interpretation of empirical $p$ values.}
\label{tab:empirical_p}
\begin{tabular}{ll}
Condition & Interpretation \\
\midrule
$\bar{p} \approx 0.63 \pm 0.03$ & Renewal-type behavior (memoryless regime) \\
$\bar{p} < 0.60$ & Overshoot persistence; event clustering \\
$\bar{p} > 0.66$ & Frequent reversals; anti-persistent dynamics \\
\end{tabular}
\end{table}

\subsection{Geometric Overshoot-Count Test}
\label{sec:geo_test}

Between two consecutive directional changes, the DcOS sequence may contain
a variable number of overshoots $K \in \{0, 1, 2, \dots\}$.  
If the process is memoryless, $K$ should follow a geometric distribution:
\begin{equation}
P(K = k) = (1 - p)^{k} p,
\label{eq:geom_pmf}
\end{equation}
with $p$ corresponding to the probability of a directional change.

The parameter $p$ was estimated as:
\begin{equation}
\hat{p}_{\mathrm{geom}} = \frac{1}{1 + \bar{K}},
\end{equation}
where $\bar{K}$ is the mean overshoot count per cycle.

Goodness-of-fit was assessed using:
\begin{itemize}
\item a \textbf{Chi-squared test}, comparing observed and expected frequencies:
\begin{equation}
\chi^2 = \sum_{k} \frac{(O_k - E_k)^2}{E_k},
\end{equation}
where $O_k$ and $E_k$ are observed and expected counts under
Eq.~\ref{eq:geom_pmf}, and
\item a \textbf{Kolmogorov--Smirnov test} using the empirical distribution of $K$.
\end{itemize}
Both tests return $p$-values ($p_{\chi^2}$, $p_{\mathrm{KS}}$)
interpreted under the null hypothesis that the overshoot counts follow
a geometric law (Table \ref{tab:geom_p}).

\begin{table}[h!]
\centering
\caption{Interpretation of geometric test results.}
\label{tab:geom_p}
\begin{tabular}{ll}
Result & Interpretation \\
\midrule
$\hat{p}_{\mathrm{geom}} \approx 0.63 \pm 0.03$ & Consistent with memoryless process \\
$\hat{p}_{\mathrm{geom}} < 0.60$ & Trending regime; overshoot persistence \\
$\hat{p}_{\mathrm{geom}} > 0.66$ & Choppy regime; frequent reversals \\
$p_{\chi^2}, p_{\mathrm{KS}} > 0.05$ & Fail to reject geometric fit (fit is geometric) \\
$p_{\chi^2}, p_{\mathrm{KS}} <= 0.05$ & Reject fit; dependence or finite-sample effect \\
\end{tabular}
\end{table}

\subsection{Exponential Overshoot-Length Test}
\label{sec:exp_test}

Each overshoot has a logarithmic length $x = |\log(S_{\max}/S_{\mathrm{Dc}})|$,
normalized by its corresponding threshold $\delta$.  
If the process is scale-invariant and memoryless, $x$ should follow an
exponential distribution:
\begin{equation}
f(x; \lambda) = \lambda e^{-\lambda x}, \qquad x > 0.
\end{equation}

The rate parameter $\lambda$ was estimated as the reciprocal of the mean:
\begin{equation}
\hat{\lambda} = \frac{1}{\bar{x}}.
\end{equation}

To test goodness of fit, a Kolmogorov--Smirnov statistic was computed
between the empirical distribution of normalized overshoot lengths
and the theoretical $\mathrm{Exp}(\hat{\lambda})$ model.
A 95\% confidence interval for $\lambda$ was derived via
normal approximation:
\begin{equation}
\hat{\lambda}_{\pm} =
\hat{\lambda} \left(1 \pm \frac{1.96}{\sqrt{n}}\right),
\end{equation}
where $n$ is the number of overshoots.

Under the exponential hazard model, the implied probability that the
next event is a directional change is:
\begin{equation}
p_{\mathrm{pred}} = 1 - e^{-\hat{\lambda}}.
\label{eq:p_pred}
\end{equation}
This prediction (Table \ref{tab:exp_p}) can be directly compared with the empirical $\bar{p}$ from
\S\ref{sec:p_mean}.

\begin{table}[h!]
\centering
\caption{Interpretation of exponential overshoot-length test.}
\label{tab:exp_p}
\begin{tabular}{ll}
Condition & Interpretation \\
\midrule
$\hat{\lambda} \approx 1$ & Ideal exponential regime (scale-invariant) \\
$\hat{\lambda} < 1$ & Longer overshoots; persistence / volatility clustering \\
$\hat{\lambda} > 1$ & Shorter overshoots; anti-persistence \\
$p_{\mathrm{KS}} > 0.05$ & Fail to reject exponential fit (fit is exponential) \\
$p_{\mathrm{KS}} <= 0.05$ & Reject exponentiality; regime shift or nonstationarity \\
\end{tabular}
\end{table}

\subsection{Integrated Consistency Checks}
\label{sec:integrated}

When the DcOS mechanism is consistent across scales, the following
relations hold simultaneously:
\begin{equation}
p_{\mathrm{mean}} \approx p_{\mathrm{geom}} \approx p_{\mathrm{pred}} \approx 1 - e^{-1},
\qquad \hat{\lambda} \approx 1.
\end{equation}
In this case, the DcOS event process behaves as a
\emph{renewal process with exponential hazard rate},
and the counts of events across thresholds $\delta$ follow
a power-law relation $N(\delta) \propto \delta^{-\beta}$ (Table \ref{tab:integrated}) consistent with fractal scaling of intrinsic time.

\begin{table}[h!]
\centering
\caption{Overall interpretation of DcOS statistical diagnostics.}
\label{tab:integrated}
\begin{tabular}{ll}
Pattern of results & Interpretation \\
\midrule
$p_{\mathrm{mean}} \simeq p_{\mathrm{geom}} \simeq p_{\mathrm{pred}} \simeq 0.63$, $\hat{\lambda}\simeq1$ &
Renewal-type intrinsic time process (scaling regime) \\
Deviations for small $\delta$ ($<10^{-3}$) & Microstructure correlations / noise \\
Deviations for large $\delta$ ($>10^{-1}$) & Undersampling of macro events \\
Stable $\lambda$ across scales & Scale invariance of overshoot distribution \\
\end{tabular}
\end{table}

Collectively, these diagnostics provide a quantitative test for the
hypothesis that the market's intrinsic time evolves through a sequence of
independent exponential waiting times between directional changes, producing
the characteristic ratio of approximately $1 - e^{-1}$ directional-change
events among all observed events.

\section{Results}
\label{sec:results}
Full results tables are shown in Appendix 1.

\subsection{Event frequency scaling law}
\label{sec:res_ev_freq}
\begin{figure}[p]
    \centering
    \includegraphics[width=\textwidth]{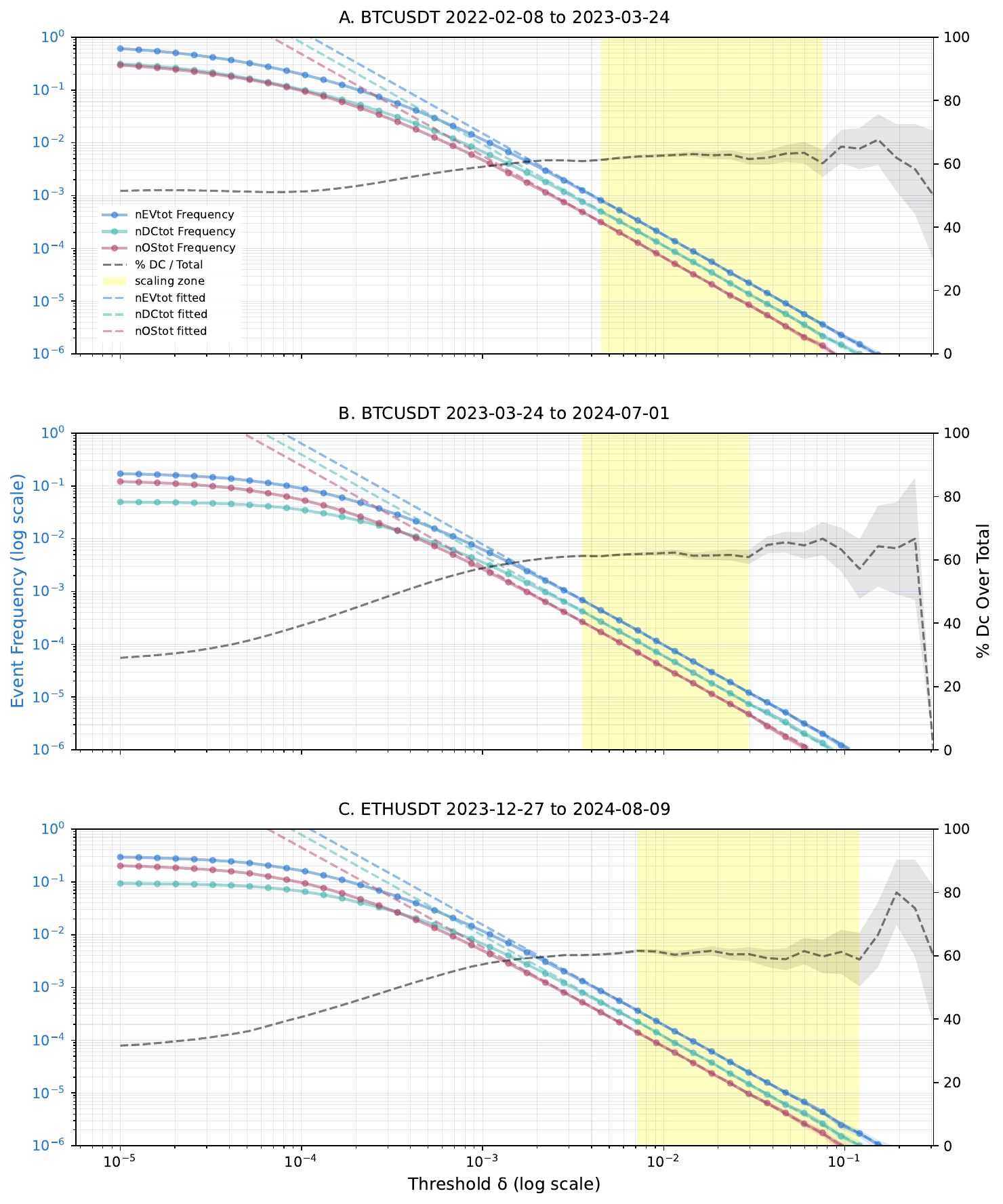}
    \caption{Event frequency per type (left y-axis) and Dc over total event frequency ratio (right y-axis) as functions of threshold $\delta$ for all three datasets. Highlighted region indicates the scaling zone (61.21$\pm$2.5\%). Within this zone, event frequency shows linear association with $\delta$ in log-log space and Dc ratio plateaus.}
    \label{fig:fig_1}
\end{figure}

Figure \ref{fig:fig_1} plots observed event frequency by type and Dc ratio. Within an intermediate $\delta$ range (e.g., \(10^{-3} < \delta < 10^{-1}\)) and across datasets, event frequency in intrinsic time obey approximate power‐law relationships across thresholds, consistent with earlier studies, e.g., \cite{glattfelder2011patterns}. We will refer to this range as the “scaling zone”. We define the scaling zone algorithmically as:
\begin{enumerate}
    \item Identify the smallest $\delta$ where $\text{dc\_pct} \ge 61.21\%$. The reasons why the \emph{empirical target} is 2\% lower than the theoretical constant are described in \S\ref{sec:deviation}.
    \item Extend to the largest consecutive $\delta$ where $\text{dc\_pct}$ remains within $61.21\% \pm 2.5\%$
    \item All smaller $\delta$ form the microstructure noise zone; larger $\delta$ form the data scarcity zone.
\end{enumerate}
Within the scaling zone, log–log plots of \(n_{\mathrm{EVtot}}\), \(n_{\mathrm{Dctot}}\), and \(n_{\mathrm{OStot}}\) versus \(\delta\) reveal linear relationships with nearly perfect fit and nearly identical slopes, indicating a self-similar regime (Table \ref{tab:regression_stats}). Regression coefficients are also remarkably stable across dataset time frames and assets, ranging from $\beta$=-1.93 (dataset A, all event types) to $\beta$=-1.89 (dataset C, OS).

In the extreme small \(\delta\) regime (e.g., \(\delta < 10^{-3}\)), event counts explode, overshoot phases dominate the cycle. Here, microstructure noise (tick size, latency, and irregular sampling)  and discretization limits (time and price granularity) inject high‑frequency mean reversion, raise event frequency toward a ceiling, and flatten the log–log curve. We refer to this regime as the “microstructure noise” zone. Conversely, in the extreme large \(\delta\) region (e.g., \(\delta > 10^{-1}\)), event counts become sparse, statistical estimation becomes unreliable, and finite-sample effects and truncation bias emerge. We refer to this regime as the “data scarcity” zone.

\begin{table}[h!]
\caption{Regression statistics for tot, dc, and os event frequencies}
\label{tab:regression_stats}
\centering
\small
\setlength{\tabcolsep}{6pt} 
\begin{tabular}{lccc ccc ccc}
 & \multicolumn{3}{c}{Tot EV} & \multicolumn{3}{c}{Dc EV} & \multicolumn{3}{c}{OS EV} \\
Dataset & $\beta$ & $R^2$ & \emph{p} & $\beta$ & $R^2$ & \emph{p} & $\beta$ & $R^2$ & \emph{p} \\
\midrule
A & -1.93 & 1.00 & \emph{p}$<$0.001\textbf{***} 
  & -1.93 & 1.00 & \emph{p}$<$0.001\textbf{***}
  & -1.93 & 1.00 & \emph{p}$<$0.001\textbf{***} \\
B & -1.91 & 1.00 & \emph{p}$<$0.001\textbf{***}
  & -1.91 & 1.00 & \emph{p}$<$0.001\textbf{***}
  & -1.91 & 1.00 & \emph{p}$<$0.001\textbf{***} \\
C & -1.90 & 1.00 & \emph{p}$<$0.001\textbf{***}
  & -1.91 & 1.00 & \emph{p}$<$0.001\textbf{***}
  & -1.89 & 1.00 & \emph{p}$<$0.001\textbf{***} \\
\end{tabular}
\end{table}

\subsection{Directional-change ratio}
\label{sec:res_dc_ratio}
In the microstructure noise zone, the Dc ratio is at or below 50\%, with large inter-dataset variability. In the data scarcity zone, estimates are highly unreliable. However, in the scaling region, Dc ratio stabilizes in the 60-63\% region. On average, Dc ratios are slightly smaller than 63.21\% (Table \ref{tab:scaling}). nevertheless, the observed values are close enough to the $1 - 1/e$ constant as to support the “exponential hazard” interpretation of intrinsic time. Reasons behind the systematic offset from the theoretical value are discussed at length in \S\ref{sec:deviation}. To accommodate for this bias, we define the scaling zone algorithmically as the $\delta$ region where the first Dc ratio crosses 61.21\% up to the last consecutive Dc ratio within the 61.21\% $\pm$ 2.5\% range. We define the microstructure noise zone as the $\delta$ region prior to the scaling zone and the data scarcity zone as the $\delta$ region after the scaling zone.

\begin{table}[h!]
\centering
\caption{Scaling region statistics}
\label{tab:scaling}
\begin{tabular}{lccccc}
\hline
Dataset & min $\delta$ & max $\delta$ & $n$ deltas & mean \%Dc & std \%Dc \\
\hline
A & 0.0045 & 0.0754 & 13 & 62.29 & 0.93 \\
B & 0.0036 & 0.0295 & 10 & 61.52 & 0.44 \\
C & 0.0072 & 0.1207 & 13 & 60.48 & 1.01 \\
\hline
\end{tabular}
\end{table}

\subsection{Renewal-process hypothesis tests}
\label{sec:res_tests}

We evaluate the three renewal-process diagnostics introduced in \S\ref{sec:methods} across the scaling zones identified in \S\ref{sec:res_dc_ratio}. Full numerical results for Datasets~A, B, and C are tabulated in Appendix (Supplementary Tables 1-6). Here we provide a qualitative synthesis of the findings.

\paragraph{Empirical Directional-change Probability (\S\ref{sec:p_mean}).}
The empirical Directional-change share $\bar{p}$ demonstrates remarkable stability within each dataset's scaling zone. For Dataset~A (BTCUSDT, 2022--2023), $\bar{p}$ varies only between $0.601$ and $0.635$ across thirteen thresholds spanning nearly two orders of magnitude, with a central tendency of $0.623 \pm 0.009$ (standard error). Dataset~B (BTCUSDT, 2023--2024) exhibits even tighter convergence, with $\bar{p}$ confined to $0.608$--$0.623$ (mean $0.615 \pm 0.004$) over ten thresholds. Dataset~C (ETHUSDT) shows marginally greater dispersion, from $0.588$ to $0.616$ (mean $0.605 \pm 0.010$), but remains stable relative to the theoretical benchmark $p_0 = 1 - e^{-1} \approx 0.6321$. The modest, systematic negative bias ($\bar{p}$ consistently $0.01$--$0.03$ below $p_0$) reflects either mild serial dependence in overshoot sequences or finite-sample attenuation, as elaborated in \S\ref{sec:deviation}. Crucially, the coefficient of variation of $\bar{p}$ within the scaling zone is under $2\%$ for all datasets, supporting the core renewal hypothesis of a constant hazard rate across scales.

\paragraph{Geometric Overshoot-Count Test (\S\ref{sec:geo_test}).}
The geometric fit parameter $\hat{p}_{\mathrm{geom}}$ mirrors $\bar{p}$ almost exactly, differing by less than $10^{-4}$ in most cases. This algebraic consistency validates the internal logic of the renewal framework: the mean overshoot count per cycle $\bar{K}$ yields a geometric parameter that reproduces the empirical event ratio. Formal goodness-of-fit tests, however, paint a more nuanced picture. In the microstructure-noise regime ($\delta < 10^{-3}$), both the chi-squared and Kolmogorov--Smirnov $p$-values are effectively zero, correctly rejecting the geometric null where market microstructure dominates. Within the scaling zone, $p$-values increase dramatically---for Dataset~B, $p_{\mathrm{KS}}$ reaches $0.69$ at $\delta = 0.0115$, and for Dataset~C, $p_{\mathrm{KS}} = 0.98$ at $\delta = 0.0596$---yet remain below conventional significance thresholds for most intermediate thresholds. This pattern indicates that while the geometric distribution captures the first moment accurately, higher-order deviations persist. The deviations are primarily driven by an excess of short overshoots (zero-count cycles) and a slightly heavier tail than predicted, consistent with volatility clustering.

\paragraph{Exponential Overshoot-Length Test (\S\ref{sec:exp_test}).}
Normalized overshoot lengths produce estimated hazard rates $\hat{\lambda}$ that cluster tightly around unity within the scaling zone. Dataset~A yields $\hat{\lambda} = 0.94 \pm 0.03$; Dataset~B gives $0.93 \pm 0.02$; and Dataset~C produces $0.92 \pm 0.03$. The slight downward bias relative to $\lambda = 1$ mirrors the negative bias in $\bar{p}$, reinforcing the interpretation of a renormalized effective hazard rate $\lambda_{\mathrm{eff}} < 1$. The predicted Directional-change probability $p_{\mathrm{pred}} = 1 - e^{-\hat{\lambda}}$ consequently averages $0.611$, $0.607$, and $0.599$ for Datasets~A, B, and C respectively, tracking the empirical $\bar{p}$ within $2\%$. Kolmogorov--Smirnov tests for exponentiality follow the same qualitative pattern as the geometric tests: $p_{\mathrm{KS}}$ near zero for $\delta < 10^{-3}$, then rising to $10^{-2}$--$10^{-1}$ in the scaling zone. Visual inspection of Q--Q plots (omitted for brevity) confirms that the exponential model fits the bulk of the distribution well, with deviations confined to the extreme tail where finite-sample effects and volatility clustering become visible.

\begin{figure}[p]
    \centering
    \includegraphics[width=\textwidth]{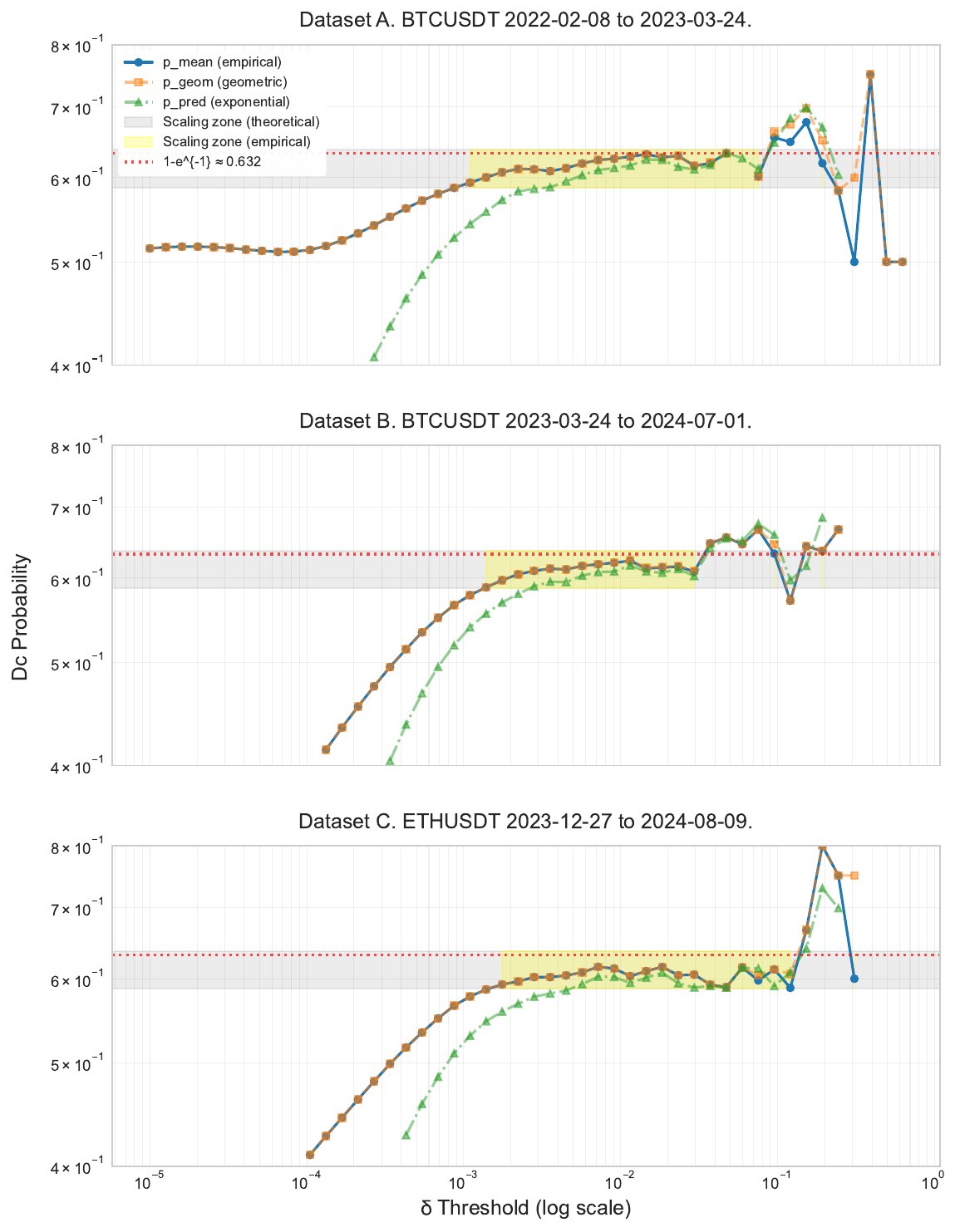}
    \caption{Comparison of three independent Directional-change probability estimates ($\bar{p}$, $\hat{p}_{\mathrm{geom}}$, and $p_{\mathrm{pred}}$) as functions of threshold $\delta$ for all three datasets. Shaded regions indicate the scaling zone (60--63\%). Within this zone, the three estimates converge to within 1--2\%, demonstrating internal consistency of the renewal model.}
    \label{fig:fig_2}
\end{figure}

\paragraph{Integrated Consistency Checks (\S\ref{sec:integrated}).}
Figure~\ref{fig:fig_2} plots the three independent estimates of the Directional-change share---$\bar{p}$ from event counts, $\hat{p}_{\mathrm{geom}}$ from overshoot-count distributions, and $p_{\mathrm{pred}}$ from exponential overshoot lengths---as functions of $\delta$. Within the scaling zone, the three curves overlap to within $1\%$--$2\%$ for all datasets, demonstrating that the renewal model's internal predictions are mutually consistent. The joint condition $\bar{p} \approx \hat{p}_{\mathrm{geom}} \approx p_{\mathrm{pred}} \approx 0.60\text{--}0.63$ is satisfied robustly for over an order of magnitude in $\delta$. Outside the scaling zone, the estimates diverge predictably: in the microstructure-noise regime, $\bar{p}$ collapses while $\hat{p}_{\mathrm{geom}}$ and $p_{\mathrm{pred}}$ remain artificially elevated due to mis-specified distributional assumptions; in the data-scarcity regime, all three become noisy as sample sizes drop below $10^3$ events. The stability of $\hat{\lambda}$ across scales (coefficient of variation $<5\%$) provides additional evidence for scale invariance of the overshoot-length distribution, a hallmark of exponential renewal.

\paragraph{Collective interpretation.}
The three diagnotic tests, while individually rejecting strict null hypotheses at conventional significance levels, collectively support the renewal-process interpretation of intrinsic time within the scaling zone. The empirical Directional-change share is stable, internally consistent, and close to the theoretical $1-e^{-1}$ benchmark; the estimated hazard rate $\hat{\lambda}$ is near unity and scale-invariant; and the three independent estimators of $p$ converge to within a few percent. The statistical rejections arise primarily from higher-order distributional mismatches---excess short cycles and modest tail heaviness---that are expected in high-frequency financial data with volatility clustering and mild serial dependence. These effects introduce a small negative bias ($\lambda_{\mathrm{eff}} \approx 0.93$) but do not invalidate the first-order exponential-hazard approximation. Consequently, the scaling zone identified by frequency scaling (\S\ref{sec:res_ev_freq}) coincides with the regime where the DcOS mechanism behaves as an approximate renewal process, providing a solid empirical foundation for the $1-e^{-1}$ heuristic.

\subsection{Summary of Findings}
\label{sec:res_summary}

\begin{itemize}
  \item \textbf{Power-law scaling}: Within the scaling zone ($10^{-3} \lesssim \delta \lesssim 10^{-1}$ in the datasets analyzed), event frequencies follow $f(\delta) \propto \delta^{-\beta}$ with $\beta \approx -1.91 \pm 0.02$ across all datasets and event types ($R^2 > 0.99$, $p < 0.001$).
  
  \item \textbf{Stable Directional-change share}: Within the scaling zone, the empirical probability $\bar{p}$ stabilizes at $0.61 \pm 0.02$, close to the theoretical $1-e^{-1} \approx 0.632$ benchmark.
  
  \item \textbf{Renewal validation}: Three independent estimators ($\bar{p}$, $\hat{p}_{\mathrm{geom}}$, $p_{\mathrm{pred}}$) converge to within 1--2\%, confirming internal consistency of the exponential-hazard model despite higher-order distributional deviations.
  
  \item \textbf{Practical heuristic}: The $1-e^{-1}$ constant reliably identifies the threshold range where scaling laws hold, providing a data-driven alternative to ad hoc $\delta$ selection.
\end{itemize}

\section{Discussion}
\label{sec:discussion}

The empirical evidence presented in \S\ref{sec:results} establishes that cryptocurrency price dynamics in Intrinsic Time exhibit two robust regularities: power-law event-frequency scaling and a stable Directional-change share near 61\%. These findings provide strong support for modeling the DcOS mechanism as a renewal process with exponential hazard. This interpretation not only rationalizes the observed scaling laws but also furnishes a practical heuristic for identifying the operational threshold range in multi-scale analyses. Below we unpack the theoretical implications, explain the systematic deviation from the idealized $1-e^{-1}$ constant, and outline directions for future research.

\subsection{Renewal Process Interpretation and the $1-e^{-1}$ Benchmark}
\label{sec:renewal_summary}

The reset property of DcOS cycles (\S\ref{subsec:renewal}) motivates a renewal-process formalism where waiting times follow $P(T \le t) = 1 - e^{-\lambda t}$. The universal constant $1-e^{-1} \approx 0.632$ emerges as the probability of a renewal within one mean period, appearing across reliability theory, queueing systems, and optimal stopping problems \cite{Balakrishnan2022Robust}.

Our empirical finding of a stable $\sim$61\% Directional-change share places market microstructure within this universality class, albeit with a renormalized effective rate $\lambda_{\mathrm{eff}} \approx 0.94$. This slight offset reflects real-market effects: volatility clustering, price discreteness, and regime mixing \cite{engle1998acd, barndorff2002econometric} (\S\ref{sec:deviation}). The $1-e^{-1}$ constant thus serves as both a theoretical anchor and a practical diagnostic: when $\bar{p} \notin [0.60,0.63]$, the chosen $\delta$ lies outside the scaling zone.

\subsection{Deviation from the Theoretical 63\% Constant}
\label{sec:deviation}

Despite the strong theoretical support, the empirical plateau sits systematically at $\sim$61\%, roughly 2--3 percentage points below $1 - e^{-1}$. This offset is not a statistical artifact but a robust feature of real, discrete markets. Several mechanisms contribute:

\begin{enumerate}
    \item \emph{Volatility clustering and weak serial dependence}: Overshoots exhibit mild persistence, inflating the mean overshoot length beyond the exponential expectation and reducing the Dc share.
    
    \item \emph{Price discreteness and bid--ask asymmetry}: Tick-size effects and asymmetric liquidity provision delay reversals, mechanically lowering the observed Dc frequency.
    
    \item \emph{Finite-threshold effects}: Logarithmic scaling introduces a small upward bias in effective trigger distance, equivalent to a renormalization of $\lambda$.
    
    \item \emph{Regime mixing}: Heterogeneous volatility states produce heavier-tailed overshoot distributions, effectively reducing the mean hazard rate to $\lambda_{\mathrm{eff}} \approx 0.94$.
\end{enumerate}

Substituting $\lambda_{\mathrm{eff}} = 0.94$ into $p = 1 - e^{-\lambda_{\mathrm{eff}}}$ yields $p \approx 0.61$, matching the empirical plateau. The ``intrinsic-time constant'' should therefore be understood as a \emph{renormalized exponential limit} characteristic of real markets rather than a failure of the renewal hypothesis. This effective constant remains stable across assets and time periods, making it a reliable diagnotic.

\subsection{Practical Applications and Future Work}
\label{sec:applications}

The $1 - e^{-1}$ heuristic offers several concrete benefits for practitioners:

\begin{itemize}
    \item \emph{Threshold selection}: The $\delta$ range where $\bar{p}$ falls within 60--63\% automatically identifies the scaling regime, eliminating ad hoc choices.
    \item \emph{Model calibration}: Fractal models of market activity can be anchored by the empirical $p$ value, ensuring consistency with microstructure dynamics.
    \item \emph{Cross-asset normalization}: The constant provides a dimensionless metric for comparing intrinsic-time dynamics across instruments, sampling frequencies, and market conditions.
\end{itemize}

Future research should extend this analysis in three directions. First, cross-asset universality must be tested across equities, commodities, and foreign exchange to confirm whether the $\sim$61\% plateau generalizes beyond cryptocurrency markets. Second, extreme-threshold behavior should be examined more closely, particularly the transition from exponential to Weibull or Gamma distributions at very large $\delta$ where finite-sample effects dominate. Third, the link between the dynamic hazard rate $\lambda$ and observable market states---such as realized volatility, bid--ask spread, or order-flow imbalance---should be investigated to develop a time-varying extension of the model.

\section{Conclusion}
\label{sec:conclusion}

This study demonstrates that Intrinsic Time behaves as an approximate renewal process governed by an exponential hazard with an effective rate $\lambda_{\mathrm{eff}} \approx 0.94$, yielding a stable Directional-change share near $1 - e^{-\lambda_{\mathrm{eff}}} \approx 0.61$. This constant not only governs the renewal of intrinsic events but also delineates the regime where power-law scaling emerges across thresholds. By connecting microstructure event statistics to fundamental principles of stochastic renewal, the $1 - e^{-1}$ heuristic offers a theoretically grounded yet practical tool for multi-scale analysis of financial markets. The empirical robustness of this constant across assets and time periods suggests that it captures a fundamental rhythm of market activity---a hidden temporal signature that bridges the gap between fractal price dynamics and the memoryless processes that generate them.

\newpage
\bibliographystyle{plain}
\bibliography{references}

@article{aloud2011directional,
  author  = {Aloud, Monira and Tsang, Edward and Olsen, Richard and Dupuis, Alexandre},
  title   = {A Directional-Change Events Approach for Studying Financial Time Series},
  journal = {Economics: The Open-Access, Open-Assessment E-Journal},
  year    = {2012},
  volume  = {6},
  number  = {2012-36},
  doi     = {10.5018/economics-ejournal.ja.2012-36},
  url     = {https://doi.org/10.5018/economics-ejournal.ja.2012-36}
}

@misc{glattfelder2025modern,
  author       = {Glattfelder, James B. and Houweling, Thomas and Olsen, Richard B.},
  title        = {A Modern Paradigm for Algorithmic Trading},
  howpublished = {arXiv preprint arXiv:2501.06032},
  year         = {2025},
  url          = {https://arxiv.org/abs/2501.06032}
}

@article{ma2017volatility,
  author  = {Ma, Junjun and Xiong, Xiong and He, Feng and Zhang, Wei},
  title   = {Volatility Measurement with Directional Change in Chinese Stock Market: Statistical Property and Investment Strategy},
  journal = {Physica A: Statistical Mechanics and its Applications},
  year    = {2017},
  volume  = {471},
  pages   = {169--180},
  doi     = {10.1016/j.physa.2016.11.113}
}

@article{petrov2019instantaneous,
  author  = {Petrov, Vladimir and Golub, Anton and Olsen, Richard B.},
  title   = {Instantaneous Volatility Seasonality of High-Frequency Markets in Directional-Change Intrinsic Time},
  journal = {Journal of Risk and Financial Management},
  year    = {2019},
  volume  = {12},
  number  = {2},
  pages   = {54},
  doi     = {10.3390/jrfm12020054}
}

@unpublished{petrov2019multidimensional,
  author = {Petrov, Vladimir and Golub, Anton and Olsen, Richard B.},
  title  = {Intrinsic Time Directional-Change Methodology in Higher Dimensions},
  note   = {SSRN Working Paper No.\ 3440628},
  year   = {2019},
  url    = {https://ssrn.com/abstract=3440628}
}

@article{petrov2020agent,
  author  = {Petrov, Vladimir and Golub, Anton and Olsen, Richard B.},
  title   = {Agent-based modelling in directional-change intrinsic time},
  journal = {Quantitative Finance},
  year    = {2020},
  volume  = {20},
  number  = {3},
  pages   = {463--482},
  doi     = {10.1080/14697688.2019.1669809}
}

@misc{glattfelder2024theory,
  author       = {Glattfelder, James B.},
  title        = {The Theory of Intrinsic Time: Market Microstructure 2.0},
  howpublished = {arXiv preprint arXiv:2406.07354},
  year         = {2024},
  url          = {https://arxiv.org/abs/2406.07354}
}

@article{guillaume1997birds,
  author  = {Guillaume, Dominique M. and Dacorogna, Michel M. and Dav{\'e}, Rakhal R. and M{\"u}ller, Ulrich A. and Olsen, Richard B. and Pictet, Olivier V.},
  title   = {From the bird's eye to the microscope: A survey of new stylized facts of the intra-daily foreign exchange markets},
  journal = {Finance and Stochastics},
  year    = {1997},
  volume  = {1},
  number  = {2},
  pages   = {95--129},
  doi     = {10.1007/s007800050018}
}

@article{glattfelder2011patterns,
  author  = {Glattfelder, James B. and Dupuis, Alexandre and Olsen, Richard B.},
  title   = {Patterns in high-frequency FX data: Discovery of 12 empirical scaling laws},
  journal = {Quantitative Finance},
  year    = {2011},
  volume  = {11},
  number  = {4},
  pages   = {599--614},
  doi     = {10.1080/14697688.2010.481632}
}

@article{Balakrishnan2022Robust,
  author  = {Balakrishnan, Narayanaswamy and Castilla, Elena and Jaenada, Mar{\'i}a and Pardo, Leandro},
  title   = {Robust inference for nondestructive one-shot device testing under step-stress model with exponential lifetimes},
  journal = {Quality and Reliability Engineering International},
  year    = {2023},
  volume  = {39},
  number  = {4},
  pages   = {1192--1222},
  doi     = {10.1002/qre.3287}
}

@book{Kleinrock1975Queueing,
  author    = {Kleinrock, Leonard},
  title     = {Queueing Systems, Volume I: Theory},
  publisher = {Wiley},
  address   = {New York},
  year      = {1975},
  isbn      = {978-0471491101}
}

@article{Peshkin2017NonExponential,
  author  = {Peshkin, Murray and Volya, Alexander and Zelevinsky, Vladimir},
  title   = {Non-exponential and oscillatory decays in quantum mechanics},
  journal = {Europhysics Letters},
  year    = {2014},
  volume  = {107},
  number  = {4},
  pages   = {40001},
  doi     = {10.1209/0295-5075/107/40001}
}

@article{Elmore1948Transient,
  author  = {Elmore, W. C.},
  title   = {The Transient Response of Damped Linear Networks with Particular Regard to Wideband Amplifiers},
  journal = {Journal of Applied Physics},
  year    = {1948},
  volume  = {19},
  number  = {1},
  pages   = {55--63},
  doi     = {10.1063/1.1697872}
}

@article{engle1998acd,
  author  = {Engle, Robert F. and Russell, Jeffrey R.},
  title   = {Autoregressive Conditional Duration: A New Model for Irregularly Spaced Transaction Data},
  journal = {Econometrica},
  year    = {1998},
  volume  = {66},
  number  = {5},
  pages   = {1127--1162},
  doi     = {10.2307/2999634}
}

@article{barndorff2002econometric,
  author  = {Barndorff-Nielsen, Ole E. and Shephard, Neil},
  title   = {Econometric analysis of realized volatility and its use in estimating stochastic volatility models},
  journal = {Journal of the Royal Statistical Society: Series B (Statistical Methodology)},
  year    = {2002},
  volume  = {64},
  number  = {2},
  pages   = {253--280},
  doi     = {10.1111/1467-9868.00336}
}

\newpage
\section*{Appendix}
\label{sec:appendix}

\begin{description}

  \item[\texttt{$\delta$}] 
  Directional-change threshold $\delta$ (relative price move).

  \item[\texttt{nDc}] 
  Number of Directional-change (Dc) events at threshold $\delta$.

  \item[\texttt{nOs}] 
  Number of Overshoot (Os) events at threshold $\delta$.

  \item[\texttt{nEv}] 
  Total number of DcOS events, defined as
  \[
    \texttt{nEv} = \texttt{nDc} + \texttt{nOs}.
  \]

  \item[\texttt{fDc}] 
  Frequency of Dc events. Computed as \texttt{nDc} divided by the total number of ticks.

  \item[\texttt{fDc\_se}] 
  Standard error of \texttt{fDc}.

  \item[\texttt{fOs}] 
  Frequency of Os events.

  \item[\texttt{fOs\_se}] 
  Standard error of \texttt{fOs}.

  \item[\texttt{fEv}] 
  Frequency of all DcOS events.

  \item[\texttt{fEv\_se}] 
  Standard error of \texttt{fEv}.

  \item[\texttt{dcPct}] 
  Percentage share of Dc events among all DcOS events, defined by
  \[
    \texttt{dcPct} = 100 \cdot \frac{\texttt{fDc}}{\texttt{fEv}}.
  \]

  \item[\texttt{seDcPct}] 
  Standard error of \texttt{dcPct}, computed using the binomial approximation.


  \item[\texttt{pMean}] 
  Average of two independent estimates of the Dc “success’’ probability.

  \item[\texttt{diff}] 
  Difference between \texttt{pMean} and the theoretical benchmark $1 - e^{-1}$.

  \item[\texttt{pGeom}] 
  Estimated geometric parameter for the overshoot–count distribution (probability of ending the overshoot).

  \item[\texttt{geoChi2p}] 
  $p$-value of the chi-squared goodness-of-fit test for the geometric distribution.

  \item[\texttt{geoKSp}] 
  $p$-value of the Kolmogorov–Smirnov test for the geometric overshoot–count distribution.

  \item[\texttt{lamHat}] 
  Estimated rate parameter $\hat{\lambda}$ of the exponential distribution fitted to overshoot lengths.

  \item[\texttt{expKSp}] 
  $p$-value of the Kolmogorov–Smirnov test for exponential overshoot-length distribution.

  \item[\texttt{lamCiLow}] 
  Lower bound of the 95\% confidence interval for $\hat{\lambda}$.

  \item[\texttt{lamCiHigh}] 
  Upper bound of the 95\% confidence interval for $\hat{\lambda}$.

  \item[\texttt{pPred}] 
  Dc success probability predicted by the exponential overshoot-length model.

\end{description}

\newpage
\begin{landscape}
\begin{table}[htbp]
\centering
\tiny
\caption*{Supplementary Table 1. Dataset A (BTCUSDT 2022-02-08 to 2023-03-24) results part 1.}
\label{ST1}
\begin{tabular}{CCCCCCCCCCCCC}
$\delta$ & nDc & nOs & nEv & fDc & fDc\_se & fOs & fOs\_se & fEv & fEv\_se & dcPct & seDcPct \\
\midrule
1.000e-05 & 11071242 & 10432797 & 21504039 & 3.130e-01 & 7.798e-05 & 2.950e-01 & 7.668e-05 & 6.080e-01 & 8.209e-05 & 51.48 & 0.01 \\
1.265e-05 & 10542518 & 9881054 & 20423572 & 2.981e-01 & 7.692e-05 & 2.794e-01 & 7.545e-05 & 5.775e-01 & 8.306e-05 & 51.62 & 0.01 \\
1.600e-05 & 9936348 & 9287831 & 19224179 & 2.809e-01 & 7.558e-05 & 2.626e-01 & 7.400e-05 & 5.436e-01 & 8.376e-05 & 51.69 & 0.01 \\
2.024e-05 & 9208242 & 8608422 & 17816664 & 2.604e-01 & 7.379e-05 & 2.434e-01 & 7.216e-05 & 5.038e-01 & 8.407e-05 & 51.68 & 0.01 \\
2.560e-05 & 8399484 & 7870248 & 16269732 & 2.375e-01 & 7.156e-05 & 2.225e-01 & 6.994e-05 & 4.600e-01 & 8.381e-05 & 51.63 & 0.01 \\
3.237e-05 & 7558686 & 7112423 & 14671109 & 2.137e-01 & 6.893e-05 & 2.011e-01 & 6.740e-05 & 4.148e-01 & 8.285e-05 & 51.52 & 0.01 \\
4.095e-05 & 6689022 & 6336288 & 13025310 & 1.891e-01 & 6.585e-05 & 1.792e-01 & 6.448e-05 & 3.683e-01 & 8.111e-05 & 51.35 & 0.01 \\
5.179e-05 & 5804362 & 5528798 & 11333160 & 1.641e-01 & 6.228e-05 & 1.563e-01 & 6.107e-05 & 3.204e-01 & 7.847e-05 & 51.22 & 0.01 \\
6.551e-05 & 4977546 & 4764538 & 9742084 & 1.407e-01 & 5.847e-05 & 1.347e-01 & 5.741e-05 & 2.755e-01 & 7.512e-05 & 51.09 & 0.02 \\
8.286e-05 & 4196536 & 4011650 & 8208186 & 1.187e-01 & 5.438e-05 & 1.134e-01 & 5.332e-05 & 2.321e-01 & 7.099e-05 & 51.13 & 0.02 \\
1.048e-04 & 3483580 & 3303823 & 6787403 & 9.850e-02 & 5.011e-05 & 9.341e-02 & 4.893e-05 & 1.919e-01 & 6.622e-05 & 51.32 & 0.02 \\
1.326e-04 & 2862914 & 2669052 & 5531966 & 8.095e-02 & 4.586e-05 & 7.547e-02 & 4.442e-05 & 1.564e-01 & 6.108e-05 & 51.75 & 0.02 \\
1.677e-04 & 2311383 & 2100492 & 4411875 & 6.535e-02 & 4.156e-05 & 5.939e-02 & 3.974e-05 & 1.247e-01 & 5.556e-05 & 52.39 & 0.02 \\
2.121e-04 & 1829091 & 1609507 & 3438598 & 5.172e-02 & 3.724e-05 & 4.551e-02 & 3.505e-05 & 9.723e-02 & 4.982e-05 & 53.19 & 0.03 \\
2.683e-04 & 1423113 & 1207114 & 2630227 & 4.024e-02 & 3.304e-05 & 3.413e-02 & 3.053e-05 & 7.437e-02 & 4.412e-05 & 54.11 & 0.03 \\
3.393e-04 & 1085375 & 883257 & 1968632 & 3.069e-02 & 2.900e-05 & 2.497e-02 & 2.624e-05 & 5.566e-02 & 3.855e-05 & 55.13 & 0.04 \\
4.292e-04 & 811599 & 633978 & 1445577 & 2.295e-02 & 2.518e-05 & 1.793e-02 & 2.231e-05 & 4.087e-02 & 3.329e-05 & 56.14 & 0.04 \\
5.429e-04 & 593644 & 446381 & 1040025 & 1.679e-02 & 2.160e-05 & 1.262e-02 & 1.877e-05 & 2.941e-02 & 2.841e-05 & 57.08 & 0.05 \\
6.866e-04 & 425762 & 309198 & 734960 & 1.204e-02 & 1.834e-05 & 8.743e-03 & 1.565e-05 & 2.078e-02 & 2.399e-05 & 57.93 & 0.06 \\
8.685e-04 & 300005 & 211250 & 511255 & 8.483e-03 & 1.542e-05 & 5.973e-03 & 1.296e-05 & 1.446e-02 & 2.007e-05 & 58.68 & 0.07 \\
1.099e-03 & 207539 & 142086 & 349625 & 5.868e-03 & 1.284e-05 & 4.017e-03 & 1.064e-05 & 9.886e-03 & 1.664e-05 & 59.36 & 0.08 \\
1.389e-03 & 141680 & 94321 & 236001 & 4.006e-03 & 1.062e-05 & 2.667e-03 & 8.672e-06 & 6.673e-03 & 1.369e-05 & 60.03 & 0.10 \\
1.758e-03 & 95896 & 62068 & 157964 & 2.711e-03 & 8.744e-06 & 1.755e-03 & 7.038e-06 & 4.466e-03 & 1.121e-05 & 60.71 & 0.12 \\
2.223e-03 & 64000 & 40681 & 104681 & 1.810e-03 & 7.147e-06 & 1.150e-03 & 5.700e-06 & 2.960e-03 & 9.135e-06 & 61.14 & 0.15 \\
2.812e-03 & 41668 & 26529 & 68197 & 1.178e-03 & 5.768e-06 & 7.501e-04 & 4.604e-06 & 1.928e-03 & 7.377e-06 & 61.10 & 0.19 \\
3.556e-03 & 26910 & 17312 & 44222 & 7.609e-04 & 4.637e-06 & 4.895e-04 & 3.719e-06 & 1.250e-03 & 5.942e-06 & 60.85 & 0.23 \\
4.498e-03 & 17576 & 11123 & 28699 & 4.970e-04 & 3.748e-06 & 3.145e-04 & 2.982e-06 & 8.115e-04 & 4.788e-06 & 61.24 & 0.29 \\
5.690e-03 & 11533 & 7107 & 18640 & 3.261e-04 & 3.036e-06 & 2.009e-04 & 2.383e-06 & 5.270e-04 & 3.859e-06 & 61.87 & 0.36 \\
7.197e-03 & 7457 & 4506 & 11963 & 2.108e-04 & 2.441e-06 & 1.274e-04 & 1.898e-06 & 3.383e-04 & 3.092e-06 & 62.33 & 0.44 \\
9.103e-03 & 4763 & 2857 & 7620 & 1.347e-04 & 1.951e-06 & 8.078e-05 & 1.511e-06 & 2.155e-04 & 2.468e-06 & 62.51 & 0.55 \\
1.151e-02 & 3043 & 1803 & 4846 & 8.604e-05 & 1.560e-06 & 5.098e-05 & 1.201e-06 & 1.370e-04 & 1.968e-06 & 62.79 & 0.69 \\
1.456e-02 & 1953 & 1142 & 3095 & 5.522e-05 & 1.250e-06 & 3.229e-05 & 9.555e-07 & 8.751e-05 & 1.573e-06 & 63.10 & 0.87 \\
1.842e-02 & 1240 & 738 & 1978 & 3.506e-05 & 9.956e-07 & 2.087e-05 & 7.681e-07 & 5.593e-05 & 1.257e-06 & 62.69 & 1.09 \\
2.330e-02 & 768 & 453 & 1221 & 2.172e-05 & 7.836e-07 & 1.281e-05 & 6.018e-07 & 3.452e-05 & 9.880e-07 & 62.90 & 1.38 \\
2.947e-02 & 485 & 303 & 788 & 1.371e-05 & 6.227e-07 & 8.567e-06 & 4.922e-07 & 2.228e-05 & 7.937e-07 & 61.55 & 1.73 \\
3.728e-02 & 312 & 192 & 504 & 8.822e-06 & 4.994e-07 & 5.429e-06 & 3.918e-07 & 1.425e-05 & 6.348e-07 & 61.90 & 2.16 \\
4.715e-02 & 203 & 118 & 321 & 5.740e-06 & 4.029e-07 & 3.336e-06 & 3.071e-07 & 9.076e-06 & 5.066e-07 & 63.24 & 2.69 \\
5.964e-02 & 127 & 73 & 200 & 3.591e-06 & 3.186e-07 & 2.064e-06 & 2.416e-07 & 5.655e-06 & 3.999e-07 & 63.50 & 3.40 \\
7.543e-02 & 77 & 51 & 128 & 2.177e-06 & 2.481e-07 & 1.442e-06 & 2.019e-07 & 3.619e-06 & 3.199e-07 & 60.16 & 4.33 \\
9.541e-02 & 53 & 28 & 81 & 1.499e-06 & 2.058e-07 & 7.917e-07 & 1.496e-07 & 2.290e-06 & 2.545e-07 & 65.43 & 5.28 \\
1.207e-01 & 35 & 19 & 54 & 9.896e-07 & 1.673e-07 & 5.372e-07 & 1.232e-07 & 1.527e-06 & 2.078e-07 & 64.81 & 6.50 \\
1.526e-01 & 23 & 11 & 34 & 6.503e-07 & 1.356e-07 & 3.110e-07 & 9.378e-08 & 9.613e-07 & 1.649e-07 & 67.65 & 8.02 \\
1.931e-01 & 13 & 8 & 21 & 3.676e-07 & 1.019e-07 & 2.262e-07 & 7.997e-08 & 5.938e-07 & 1.296e-07 & 61.90 & 10.60 \\
2.442e-01 & 7 & 5 & 12 & 1.979e-07 & 7.481e-08 & 1.414e-07 & 6.322e-08 & 3.393e-07 & 9.795e-08 & 58.33 & 14.23 \\
3.089e-01 & 3 & 3 & 6 & 8.482e-08 & 4.897e-08 & 8.482e-08 & 4.897e-08 & 1.696e-07 & 6.926e-08 & 50.00 & 20.41 \\
3.907e-01 & 3 & 1 & 4 & 8.482e-08 & 4.897e-08 & 2.827e-08 & 2.827e-08 & 1.131e-07 & 5.655e-08 & 75.00 & 21.65 \\
4.942e-01 & 1 & 1 & 2 & 2.827e-08 & 2.827e-08 & 2.827e-08 & 2.827e-08 & 5.655e-08 & 3.999e-08 & 50.00 & 35.36 \\
6.251e-01 & 1 & 1 & 2 & 2.827e-08 & 2.827e-08 & 2.827e-08 & 2.827e-08 & 5.655e-08 & 3.999e-08 & 50.00 & 35.36 \\
7.906e-01 & 1 &  & 1 & 2.827e-08 & 2.827e-08 &  &  & 2.827e-08 & 2.827e-08 & 100.00 &  \\
1.000e+00 &  &  &  &  &  &  &  &  &  &  &  \\
\end{tabular}
\end{table}
\end{landscape}

\begin{landscape}
\begin{table}[htbp]
\centering
\tiny
\caption*{Supplementary Table 2. Dataset A (BTCUSDT 2022-02-08 to 2023-03-24) results part 2.}
\label{ST2}
\begin{tabular}{CCCccCcCCCC}
$\delta$ & pMean & diff & pGeom & geoChi2p & geoKSp & lamHat & expKSp & lamCiLow & lamCiHigh & pPred \\
\midrule
1.000e-05   & 0.51 & -0.12 & 0.51 & \textless{}0.0001 & \textless{}0.0001 & 0.06 & \textless{}0.0001 & 0.06 & 0.06 & 0.06 \\
1.265e-05  & 0.52 & -0.12 & 0.52 & \textless{}0.0001 & \textless{}0.0001 & 0.08 & \textless{}0.0001 & 0.08 & 0.08 & 0.08 \\
1.600e-05  & 0.52 & -0.12 & 0.52 & \textless{}0.0001 & \textless{}0.0001 & 0.10 & \textless{}0.0001 & 0.10 & 0.10 & 0.09 \\
2.024e-05  & 0.52 & -0.12 & 0.52 & \textless{}0.0001 & \textless{}0.0001 & 0.12 & \textless{}0.0001 & 0.12 & 0.12 & 0.11 \\
2.560e-05  & 0.52 & -0.12 & 0.52 & \textless{}0.0001 & \textless{}0.0001 & 0.14 & \textless{}0.0001 & 0.14 & 0.14 & 0.13 \\
3.237e-05  & 0.52 & -0.12 & 0.52 & \textless{}0.0001 & \textless{}0.0001 & 0.16 & \textless{}0.0001 & 0.16 & 0.16 & 0.15 \\
4.095e-05  & 0.51 & -0.12 & 0.51 & \textless{}0.0001 & \textless{}0.0001 & 0.19 & \textless{}0.0001 & 0.19 & 0.19 & 0.18 \\
5.179e-05  & 0.51 & -0.12 & 0.51 & \textless{}0.0001 & \textless{}0.0001 & 0.22 & \textless{}0.0001 & 0.22 & 0.23 & 0.20 \\
6.551e-05  & 0.51 & -0.12 & 0.51 & \textless{}0.0001 & \textless{}0.0001 & 0.26 & \textless{}0.0001 & 0.26 & 0.26 & 0.23 \\
8.286e-05  & 0.51 & -0.12 & 0.51 & \textless{}0.0001 & \textless{}0.0001 & 0.30 & \textless{}0.0001 & 0.30 & 0.30 & 0.26 \\
1.048e-04  & 0.51 & -0.12 & 0.51 & \textless{}0.0001 & \textless{}0.0001 & 0.34 & \textless{}0.0001 & 0.34 & 0.34 & 0.29 \\
1.326e-04  & 0.52 & -0.11 & 0.52 & \textless{}0.0001 & \textless{}0.0001 & 0.38 & \textless{}0.0001 & 0.38 & 0.38 & 0.32 \\
1.677e-04  & 0.52 & -0.11 & 0.52 & \textless{}0.0001 & \textless{}0.0001 & 0.43 & \textless{}0.0001 & 0.43 & 0.43 & 0.35 \\
2.121e-04  & 0.53 & -0.10 & 0.53 & \textless{}0.0001 & \textless{}0.0001 & 0.47 & \textless{}0.0001 & 0.47 & 0.48 & 0.38 \\
2.683e-04  & 0.54 & -0.09 & 0.54 & \textless{}0.0001 & \textless{}0.0001 & 0.52 & \textless{}0.0001 & 0.52 & 0.52 & 0.41 \\
3.393e-04  & 0.55 & -0.08 & 0.55 & \textless{}0.0001 & \textless{}0.0001 & 0.57 & \textless{}0.0001 & 0.57 & 0.57 & 0.44 \\
4.292e-04  & 0.56 & -0.07 & 0.56 & \textless{}0.0001 & \textless{}0.0001 & 0.62 & \textless{}0.0001 & 0.62 & 0.62 & 0.46 \\
5.429e-04  & 0.57 & -0.06 & 0.57 & \textless{}0.0001 & \textless{}0.0001 & 0.67 & \textless{}0.0001 & 0.66 & 0.67 & 0.49 \\
6.866e-04  & 0.58 & -0.05 & 0.58 & \textless{}0.0001 & \textless{}0.0001 & 0.71 & \textless{}0.0001 & 0.71 & 0.71 & 0.51 \\
8.685e-04  & 0.59 & -0.05 & 0.59 & \textless{}0.0001 & \textless{}0.0001 & 0.75 & \textless{}0.0001 & 0.75 & 0.75 & 0.53 \\
1.099e-03 & 0.59 & -0.04 & 0.59 & \textless{}0.0001 & \textless{}0.0001 & 0.78 & \textless{}0.0001 & 0.78 & 0.79 & 0.54 \\
1.389e-03 & 0.60 & -0.03 & 0.60 & \textless{}0.0001 & \textless{}0.0001 & 0.81 & \textless{}0.0001 & 0.81 & 0.82 & 0.56 \\
1.758e-03 & 0.61 & -0.03 & 0.61 & \textless{}0.0001 & \textless{}0.0001 & 0.85 & 0.04 & 0.84 & 0.85 & 0.57 \\
2.223e-03 & 0.61 & -0.02 & 0.61 & \textless{}0.0001 & \textless{}0.0001 & 0.87 & 0.07 & 0.87 & 0.88 & 0.58 \\
2.812e-03 & 0.61 & -0.02 & 0.61 & \textless{}0.0001 & \textless{}0.0001 & 0.88 & 0.48 & 0.87 & 0.89 & 0.59 \\
3.556e-03 & 0.61 & -0.02 & 0.61 & \textless{}0.0001 & \textless{}0.0001 & 0.89 & 0.10 & 0.88 & 0.90 & 0.59 \\
4.498e-03 & 0.61 & -0.02 & 0.61 & \textless{}0.0001 & \textless{}0.0001 & 0.90 & 0.11 & 0.89 & 0.92 & 0.60 \\
5.690e-03 & 0.62 & -0.01 & 0.62 & \textless{}0.0001 & \textless{}0.0001 & 0.92 & 0.66 & 0.91 & 0.94 & 0.60 \\
7.197e-03 & 0.62 & -0.01 & 0.62 & \textless{}0.0001 & \textless{}0.0001 & 0.94 & 0.60 & 0.92 & 0.96 & 0.61 \\
9.103e-03 & 0.63 & -0.01 & 0.63 & \textless{}0.0001 & \textless{}0.0001 & 0.95 & 0.75 & 0.92 & 0.98 & 0.61 \\
1.151e-02 & 0.63 & -0.00 & 0.63 & \textless{}0.0001 & \textless{}0.0001 & 0.96 & 0.69 & 0.92 & 0.99 & 0.62 \\
1.456e-02 & 0.63 & -0.00 & 0.63 & \textless{}0.0001 & \textless{}0.0001 & 0.98 & 0.83 & 0.93 & 1.02 & 0.62 \\
1.842e-02 & 0.63 & -0.01 & 0.63 & \textless{}0.0001 & \textless{}0.0001 & 0.98 & 0.87 & 0.92 & 1.03 & 0.62 \\
2.330e-02 & 0.63 & -0.00 & 0.63 & \textless{}0.0001 & \textless{}0.0001 & 0.95 & 0.95 & 0.88 & 1.02 & 0.61 \\
2.947e-02 & 0.62 & -0.02 & 0.62 & \textless{}0.0001 & \textless{}0.0001 & 0.94 & 0.68 & 0.86 & 1.03 & 0.61 \\
3.728e-02 & 0.62 & -0.01 & 0.62 & \textless{}0.0001 & \textless{}0.0001 & 0.96 & 0.88 & 0.85 & 1.07 & 0.62 \\
4.715e-02 & 0.63 & 0.00  & 0.63 & \textless{}0.0001 & \textless{}0.0001 & 1.00 & 0.81 & 0.86 & 1.14 & 0.63 \\
5.964e-02 & 0.64 & 0.00 & 0.64 & \textless{}0.0001 & \textless{}0.0001 & 0.98 & 0.93 & 0.81 & 1.15 & 0.63 \\
7.543e-02 & 0.60 & -0.03 & 0.60 & \textless{}0.0001 & \textless{}0.0001 & 0.94 & 0.99 & 0.73 & 1.16 & 0.61 \\
9.541e-02 & 0.65 & 0.02 & 0.66 & \textless{}0.0001 & \textless{}0.0001 & 1.04 & 0.98 & 0.76 & 1.32 & 0.65 \\
1.207e-01 & 0.65 & 0.02 & 0.67 & \textless{}0.0001 & \textless{}0.0001 & 1.15 & 0.98 & 0.77 & 1.53 & 0.68 \\
1.526e-01 & 0.68 & 0.04 & 0.70 & \textless{}0.0001 & \textless{}0.0001 & 1.20 & 0.90 & 0.71 & 1.69 & 0.70 \\
1.931e-01 & 0.62 & -0.01 & 0.65 & \textless{}0.0001 & 0.03 & 1.10 & 0.93 & 0.50 & 1.71 & 0.67 \\
2.442e-01 & 0.58 & -0.05 & 0.58 & \textless{}0.0001 & 0.01 & 0.93 & 0.96 & 0.24 & 1.61 & 0.60 \\
3.089e-01 & 0.50 & -0.13 & 0.60 & \textless{}0.0001 & 0.07 &  &  &  &  &  \\
3.907e-01 & 0.75 & 0.12 & 0.75 & \textless{}0.0001 & 0.07 &  &  &  &  &  \\
4.942e-01 & 0.50 & -0.13 & 0.50 &  &  &  &  &  &  &  \\
6.251e-01 & 0.50 & -0.13 & 0.50 &  &  &  &  &  &  &  \\
7.906e-01 & 1.00 & 0.37 & 1.00 &  &  &  &  &  &  &  \\
1.000e+00 &  &  &  &  &  &  &  &  &  &  \\
\end{tabular}
\end{table}
\end{landscape}

\newpage
\begin{landscape}
\begin{table}[htbp]
\centering
\tiny
\caption*{Supplementary Table 3. Dataset B ( BTCUSDT 2023-03-24 to 2024-07-01) results part 1.}
\label{ST3}
\begin{tabular}{CCCCCCCCCCCC}
$\delta$ & nDc & nOs & nEv & fDc & seDc & fOs & seOs & fEv & seEv & dcPct & seDcPct \\
\midrule
1.000e-05 & 1990932 & 4858319 & 6849251 & 4.961e-02 & 3.428e-05 & 1.211e-01 & 5.149e-05 & 1.707e-01 & 5.939e-05 & 29.07 & 0.02 \\
1.265e-05 & 1979430 & 4741064 & 6720494 & 4.932e-02 & 3.418e-05 & 1.181e-01 & 5.095e-05 & 1.675e-01 & 5.894e-05 & 29.45 & 0.02 \\
1.600e-05 & 1966428 & 4606215 & 6572643 & 4.900e-02 & 3.407e-05 & 1.148e-01 & 5.032e-05 & 1.638e-01 & 5.842e-05 & 29.92 & 0.02 \\
2.024e-05 & 1947040 & 4434451 & 6381491 & 4.851e-02 & 3.391e-05 & 1.105e-01 & 4.949e-05 & 1.590e-01 & 5.772e-05 & 30.51 & 0.02 \\
2.560e-05 & 1920860 & 4232590 & 6153450 & 4.786e-02 & 3.370e-05 & 1.055e-01 & 4.848e-05 & 1.533e-01 & 5.687e-05 & 31.22 & 0.02 \\
3.237e-05 & 1882194 & 3967795 & 5849989 & 4.690e-02 & 3.337e-05 & 9.887e-02 & 4.712e-05 & 1.458e-01 & 5.570e-05 & 32.17 & 0.02 \\
4.095e-05 & 1833168 & 3681655 & 5514823 & 4.568e-02 & 3.296e-05 & 9.174e-02 & 4.556e-05 & 1.374e-01 & 5.435e-05 & 33.24 & 0.02 \\
5.179e-05 & 1752306 & 3313459 & 5065765 & 4.366e-02 & 3.226e-05 & 8.256e-02 & 4.344e-05 & 1.262e-01 & 5.242e-05 & 34.59 & 0.02 \\
6.551e-05 & 1653186 & 2918535 & 4571721 & 4.119e-02 & 3.137e-05 & 7.272e-02 & 4.099e-05 & 1.139e-01 & 5.015e-05 & 36.16 & 0.02 \\
8.286e-05 & 1538022 & 2525169 & 4063191 & 3.832e-02 & 3.030e-05 & 6.292e-02 & 3.833e-05 & 1.012e-01 & 4.762e-05 & 37.85 & 0.02 \\
1.048e-04 & 1386102 & 2113198 & 3499300 & 3.454e-02 & 2.882e-05 & 5.266e-02 & 3.526e-05 & 8.719e-02 & 4.453e-05 & 39.61 & 0.03 \\
1.326e-04 & 1214246 & 1718682 & 2932928 & 3.026e-02 & 2.704e-05 & 4.282e-02 & 3.196e-05 & 7.308e-02 & 4.108e-05 & 41.40 & 0.03 \\
1.677e-04 & 1044456 & 1360467 & 2404923 & 2.602e-02 & 2.513e-05 & 3.390e-02 & 2.857e-05 & 5.992e-02 & 3.747e-05 & 43.43 & 0.04 \\
2.121e-04 & 872978 & 1048029 & 1921007 & 2.175e-02 & 2.303e-05 & 2.611e-02 & 2.517e-05 & 4.787e-02 & 3.370e-05 & 45.44 & 0.04 \\
2.683e-04 & 713508 & 789410 & 1502918 & 1.778e-02 & 2.086e-05 & 1.967e-02 & 2.192e-05 & 3.745e-02 & 2.997e-05 & 47.47 & 0.04 \\
3.393e-04 & 566560 & 577781 & 1144341 & 1.412e-02 & 1.862e-05 & 1.440e-02 & 1.880e-05 & 2.851e-02 & 2.627e-05 & 49.51 & 0.05 \\
4.292e-04 & 439751 & 414837 & 854588 & 1.096e-02 & 1.643e-05 & 1.034e-02 & 1.597e-05 & 2.129e-02 & 2.279e-05 & 51.46 & 0.05 \\
5.429e-04 & 332895 & 290895 & 623790 & 8.295e-03 & 1.432e-05 & 7.248e-03 & 1.339e-05 & 1.554e-02 & 1.953e-05 & 53.37 & 0.06 \\
6.866e-04 & 245857 & 200603 & 446460 & 6.126e-03 & 1.232e-05 & 4.998e-03 & 1.113e-05 & 1.112e-02 & 1.656e-05 & 55.07 & 0.07 \\
8.685e-04 & 177053 & 135725 & 312778 & 4.412e-03 & 1.046e-05 & 3.382e-03 & 9.164e-06 & 7.794e-03 & 1.388e-05 & 56.61 & 0.09 \\
1.099e-03 & 124821 & 90963 & 215784 & 3.110e-03 & 8.790e-06 & 2.267e-03 & 7.507e-06 & 5.377e-03 & 1.154e-05 & 57.85 & 0.11 \\
1.389e-03 & 85926 & 60172 & 146098 & 2.141e-03 & 7.296e-06 & 1.499e-03 & 6.108e-06 & 3.640e-03 & 9.507e-06 & 58.81 & 0.13 \\
1.758e-03 & 58402 & 39363 & 97765 & 1.455e-03 & 6.017e-06 & 9.808e-04 & 4.941e-06 & 2.436e-03 & 7.781e-06 & 59.74 & 0.16 \\
2.223e-03 & 39095 & 25529 & 64624 & 9.741e-04 & 4.924e-06 & 6.361e-04 & 3.980e-06 & 1.610e-03 & 6.329e-06 & 60.50 & 0.19 \\
2.812e-03 & 25851 & 16557 & 42408 & 6.441e-04 & 4.005e-06 & 4.126e-04 & 3.206e-06 & 1.057e-03 & 5.129e-06 & 60.96 & 0.24 \\
3.556e-03 & 16877 & 10681 & 27558 & 4.205e-04 & 3.236e-06 & 2.661e-04 & 2.575e-06 & 6.867e-04 & 4.135e-06 & 61.24 & 0.29 \\
4.498e-03 & 10793 & 6858 & 17651 & 2.689e-04 & 2.588e-06 & 1.709e-04 & 2.063e-06 & 4.398e-04 & 3.310e-06 & 61.15 & 0.37 \\
5.690e-03 & 7045 & 4389 & 11434 & 1.755e-04 & 2.091e-06 & 1.094e-04 & 1.651e-06 & 2.849e-04 & 2.664e-06 & 61.61 & 0.45 \\
7.197e-03 & 4547 & 2807 & 7354 & 1.133e-04 & 1.680e-06 & 6.994e-05 & 1.320e-06 & 1.832e-04 & 2.137e-06 & 61.83 & 0.57 \\
9.103e-03 & 2887 & 1767 & 4654 & 7.194e-05 & 1.339e-06 & 4.403e-05 & 1.047e-06 & 1.160e-04 & 1.700e-06 & 62.03 & 0.71 \\
1.151e-02 & 1859 & 1125 & 2984 & 4.632e-05 & 1.074e-06 & 2.803e-05 & 8.357e-07 & 7.435e-05 & 1.361e-06 & 62.30 & 0.89 \\
1.456e-02 & 1163 & 735 & 1898 & 2.898e-05 & 8.497e-07 & 1.831e-05 & 6.755e-07 & 4.729e-05 & 1.086e-06 & 61.28 & 1.12 \\
1.842e-02 & 733 & 461 & 1194 & 1.826e-05 & 6.746e-07 & 1.149e-05 & 5.350e-07 & 2.975e-05 & 8.610e-07 & 61.39 & 1.41 \\
2.330e-02 & 475 & 297 & 772 & 1.184e-05 & 5.431e-07 & 7.400e-06 & 4.294e-07 & 1.924e-05 & 6.923e-07 & 61.53 & 1.75 \\
2.947e-02 & 297 & 191 & 488 & 7.400e-06 & 4.294e-07 & 4.759e-06 & 3.444e-07 & 1.216e-05 & 5.504e-07 & 60.86 & 2.21 \\
3.728e-02 & 207 & 113 & 320 & 5.158e-06 & 3.585e-07 & 2.816e-06 & 2.649e-07 & 7.974e-06 & 4.457e-07 & 64.69 & 2.67 \\
4.715e-02 & 135 & 71 & 206 & 3.364e-06 & 2.895e-07 & 1.769e-06 & 2.100e-07 & 5.133e-06 & 3.576e-07 & 65.53 & 3.31 \\
5.964e-02 & 82 & 45 & 127 & 2.043e-06 & 2.256e-07 & 1.121e-06 & 1.672e-07 & 3.164e-06 & 2.808e-07 & 64.57 & 4.24 \\
7.543e-02 & 54 & 27 & 81 & 1.346e-06 & 1.831e-07 & 6.728e-07 & 1.295e-07 & 2.018e-06 & 2.243e-07 & 66.67 & 5.24 \\
9.541e-02 & 31 & 18 & 49 & 7.724e-07 & 1.387e-07 & 4.485e-07 & 1.057e-07 & 1.221e-06 & 1.744e-07 & 63.27 & 6.89 \\
1.207e-01 & 16 & 12 & 28 & 3.987e-07 & 9.967e-08 & 2.990e-07 & 8.632e-08 & 6.977e-07 & 1.318e-07 & 57.14 & 9.35 \\
1.526e-01 & 9 & 5 & 14 & 2.243e-07 & 7.475e-08 & 1.246e-07 & 5.572e-08 & 3.488e-07 & 9.323e-08 & 64.29 & 12.81 \\
1.931e-01 & 7 & 4 & 11 & 1.744e-07 & 6.592e-08 & 9.967e-08 & 4.983e-08 & 2.741e-07 & 8.264e-08 & 63.64 & 14.50 \\
2.442e-01 & 4 & 2 & 6 & 9.967e-08 & 4.983e-08 & 4.983e-08 & 3.524e-08 & 1.495e-07 & 6.103e-08 & 66.67 & 19.25 \\
3.089e-01 & 0 & 2 & 2 &  &  & 4.983e-08 & 3.524e-08 & 4.983e-08 & 3.524e-08 &  &  \\
3.907e-01 & 0 & 1 & 1 &  &  & 2.492e-08 & 2.492e-08 & 2.492e-08 & 2.492e-08 &  &  \\
4.942e-01 & 0 & 1 & 1 &  &  & 2.492e-08 & 2.492e-08 & 2.492e-08 & 2.492e-08 &  &  \\
6.251e-01 & 0 & 0 & 0 &  &  &  &  &  &  &  &  \\
7.906e-01 & 0 & 0 & 0 &  &  &  &  &  &  &  &  \\
1.000e+00 & 0 & 0 & 0 &  &  &  &  &  &  &  &  \\

\end{tabular}
\end{table}
\end{landscape}

\newpage
\begin{landscape}
\begin{table}[htbp]
\centering
\tiny
\caption*{Supplementary Table 4. Dataset B ( BTCUSDT 2023-03-24 to 2024-07-01) results part 1.}
\label{ST4}
\begin{tabular}{CCCCCCCCCCC}
$\delta$ & pMean & diff & pGeom & geoChi2p & geoKSp & lamHat & expKSp & lamCiLow & lamCiHigh & pPred \\
\midrule
1.000e-05 & 0.29 & -0.34 & 0.29 & \textless{}0.0001 & \textless{}0.0001 & 0.03 & \textless{}0.0001 & 0.03 & 0.03 & 0.03 \\
1.265e-05 & 0.29 & -0.34 & 0.29 & \textless{}0.0001 & \textless{}0.0001 & 0.04 & \textless{}0.0001 & 0.04 & 0.04 & 0.04 \\
1.600e-05 & 0.30 & -0.33 & 0.30 & \textless{}0.0001 & \textless{}0.0001 & 0.05 & \textless{}0.0001 & 0.05 & 0.05 & 0.05 \\
2.024e-05 & 0.31 & -0.33 & 0.31 & \textless{}0.0001 & \textless{}0.0001 & 0.06 & \textless{}0.0001 & 0.06 & 0.06 & 0.06 \\
2.560e-05 & 0.31 & -0.32 & 0.31 & \textless{}0.0001 & \textless{}0.0001 & 0.07 & \textless{}0.0001 & 0.07 & 0.07 & 0.07 \\
3.237e-05 & 0.32 & -0.31 & 0.32 & \textless{}0.0001 & \textless{}0.0001 & 0.09 & \textless{}0.0001 & 0.09 & 0.09 & 0.09 \\
4.095e-05 & 0.33 & -0.30 & 0.33 & \textless{}0.0001 & \textless{}0.0001 & 0.12 & \textless{}0.0001 & 0.12 & 0.12 & 0.11 \\
5.179e-05 & 0.35 & -0.29 & 0.35 & \textless{}0.0001 & \textless{}0.0001 & 0.14 & \textless{}0.0001 & 0.14 & 0.14 & 0.13 \\
6.551e-05 & 0.36 & -0.27 & 0.36 & \textless{}0.0001 & \textless{}0.0001 & 0.18 & \textless{}0.0001 & 0.18 & 0.18 & 0.16 \\
8.286e-05 & 0.38 & -0.25 & 0.38 & \textless{}0.0001 & \textless{}0.0001 & 0.21 & \textless{}0.0001 & 0.21 & 0.21 & 0.19 \\
1.048e-04 & 0.40 & -0.24 & 0.40 & \textless{}0.0001 & \textless{}0.0001 & 0.26 & \textless{}0.0001 & 0.26 & 0.26 & 0.23 \\
1.326e-04 & 0.41 & -0.22 & 0.41 & \textless{}0.0001 & \textless{}0.0001 & 0.30 & \textless{}0.0001 & 0.30 & 0.30 & 0.26 \\
1.677e-04 & 0.43 & -0.20 & 0.43 & \textless{}0.0001 & \textless{}0.0001 & 0.35 & \textless{}0.0001 & 0.35 & 0.35 & 0.30 \\
2.121e-04 & 0.45 & -0.18 & 0.45 & \textless{}0.0001 & \textless{}0.0001 & 0.41 & \textless{}0.0001 & 0.40 & 0.41 & 0.33 \\
2.683e-04 & 0.47 & -0.16 & 0.47 & \textless{}0.0001 & \textless{}0.0001 & 0.46 & \textless{}0.0001 & 0.46 & 0.46 & 0.37 \\
3.393e-04 & 0.50 & -0.14 & 0.50 & \textless{}0.0001 & \textless{}0.0001 & 0.52 & \textless{}0.0001 & 0.52 & 0.52 & 0.40 \\
4.293e-04 & 0.51 & -0.12 & 0.51 & \textless{}0.0001 & \textless{}0.0001 & 0.57 & \textless{}0.0001 & 0.57 & 0.58 & 0.44 \\
5.429e-04 & 0.53 & -0.10 & 0.53 & \textless{}0.0001 & \textless{}0.0001 & 0.63 & \textless{}0.0001 & 0.63 & 0.63 & 0.47 \\
6.866e-04 & 0.55 & -0.08 & 0.55 & \textless{}0.0001 & \textless{}0.0001 & 0.68 & \textless{}0.0001 & 0.66 & 0.69 & 0.50 \\
8.685e-04 & 0.57 & -0.07 & 0.57 & \textless{}0.0001 & \textless{}0.0001 & 0.73 & \textless{}0.0001 & 0.73 & 0.74 & 0.52 \\
1.099e-03 & 0.58 & -0.05 & 0.58 & \textless{}0.0001 & \textless{}0.0001 & 0.76 & \textless{}0.0001 & 0.75 & 0.76 & 0.53 \\
1.389e-03 & 0.59 & -0.04 & 0.59 & \textless{}0.0001 & \textless{}0.0001 & 0.79 & \textless{}0.01 & 0.79 & 0.80 & 0.55 \\
1.758e-03 & 0.60 & -0.03 & 0.60 & \textless{}0.0001 & \textless{}0.0001 & 0.84 & \textless{}0.05 & 0.84 & 0.85 & 0.56 \\
2.223e-03 & 0.61 & -0.03 & 0.61 & \textless{}0.0001 & \textless{}0.0001 & 0.87 & \textless{}0.05 & 0.87 & 0.88 & 0.58 \\
2.812e-03 & 0.61 & -0.02 & 0.61 & \textless{}0.0001 & \textless{}0.0001 & 0.89 & 0.17 & 0.88 & 0.90 & 0.59 \\
3.556e-03 & 0.61 & -0.02 & 0.61 & \textless{}0.0001 & \textless{}0.05 & 0.87 & 0.58 & 0.86 & 0.88 & 0.59 \\
4.498e-03 & 0.61 & -0.02 & 0.61 & \textless{}0.0001 & 0.92 & 0.90 & 0.92 & 0.89 & 0.92 & 0.60 \\
5.690e-03 & 0.62 & -0.02 & 0.62 & \textless{}0.0001 & 0.95 & 0.92 & 0.95 & 0.90 & 0.95 & 0.60 \\
7.197e-03 & 0.62 & -0.01 & 0.62 & \textless{}0.0001 & 0.84 & 0.94 & 0.84 & 0.91 & 0.96 & 0.61 \\
9.103e-03 & 0.62 & -0.01 & 0.62 & \textless{}0.0001 & 0.55 & 0.94 & 0.55 & 0.90 & 0.97 & 0.61 \\
1.151e-02 & 0.62 & -0.01 & 0.62 & \textless{}0.0001 & 0.95 & 0.96 & 0.95 & 0.92 & 1.00 & 0.62 \\
1.456e-02 & 0.61 & -0.02 & 0.61 & \textless{}0.0001 & 0.97 & 0.94 & 0.97 & 0.88 & 0.99 & 0.61 \\
1.842e-02 & 0.61 & -0.02 & 0.61 & \textless{}0.0001 & \textless{}0.0001 & 0.93 & 0.94 & 0.87 & 1.00 & 0.61 \\
2.330e-02 & 0.62 & -0.02 & 0.62 & \textless{}0.0001 & \textless{}0.0001 & 0.95 & 0.97 & 0.86 & 1.03 & 0.61 \\
2.947e-02 & 0.61 & -0.02 & 0.61 & \textless{}0.0001 & \textless{}0.0001 & 0.92 & 0.49 & 0.82 & 1.03 & 0.60 \\
3.728e-02 & 0.65 & 0.01 & 0.65 & \textless{}0.0001 & \textless{}0.001 & 1.02 & 0.77 & 0.88 & 1.16 & 0.64 \\
4.715e-02 & 0.66 & 0.02 & 0.66 & \textless{}0.0001 & \textless{}0.001 & 1.06 & 0.98 & 0.88 & 1.24 & 0.65 \\
5.964e-02 & 0.65 & 0.01 & 0.65 & \textless{}0.0001 & \textless{}0.001 & 1.05 & 0.98 & 0.82 & 1.28 & 0.65 \\
7.543e-02 & 0.67 & 0.03 & 0.67 & \textless{}0.0001 & \textless{}0.001 & 1.12 & 0.89 & 0.82 & 1.42 & 0.68 \\
9.541e-02 & 0.63 & 0.00 & 0.65 & \textless{}0.0001 & \textless{}0.001 & 1.08 & 0.46 & 0.70 & 1.45 & 0.66 \\
1.207e-01 & 0.57 & -0.06 & 0.57 & \textless{}0.0001 & \textless{}0.001 & 0.91 & 0.38 & 0.46 & 1.36 & 0.60 \\
1.526e-01 & 0.64 & 0.01 & 0.64 & \textless{}0.0001 & \textless{}0.001 & 0.96 & 0.52 & 0.33 & 1.58 & 0.62 \\
1.931e-01 & 0.64 & 0.00 & 0.64 & \textless{}0.0001 & \textless{}0.001 & 1.15 & 0.37 & 0.30 & 2.00 & 0.68 \\
2.442e-01 & 0.67 & 0.03 & 0.67 & \textless{}0.0001 & 0.19 &  &  &  &  &  \\
3.089e-01 &  & -0.63 &  &  &  &  &  &  &  &  \\
3.907e-01 &  & -0.63 &  &  &  &  &  &  &  &  \\
4.942e-01 &  & -0.63 &  &  &  &  &  &  &  &  \\
6.251e-01 &  &  &  &  &  &  &  &  &  &  \\
7.906e-01 &  &  &  &  &  &  &  &  &  &  \\
1.000e+00 &  &  &  &  &  &  &  &  &  &  \\
\end{tabular}
\end{table}
\end{landscape}

\newpage
\begin{landscape}
\begin{table}[htbp]
\centering
\tiny
\caption*{Supplementary Table 5. Dataset C (ETHUSDT 2023-12-27 to 2024-08-09) results part 1.}
\label{ST5}
\begin{tabular}{CCCCCCCCCCCC}
$\delta$ & nDc & nOs & nEv & fDc & seDc & fOs & seOs & fEv & seEv & dcPct & seDcPct \\
\midrule

1.000e-05 & 1845295 & 3985168 & 5830463 & 9.320e-02 & 6.533e-05 & 2.013e-01 & 9.011e-05 & 2.945e-01 & 1.024e-04 & 31.65 & 0.02 \\
1.265e-05 & 1835745 & 3911917 & 5747662 & 9.271e-02 & 6.518e-05 & 1.976e-01 & 8.948e-05 & 2.903e-01 & 1.020e-04 & 31.94 & 0.02 \\
1.600e-05 & 1821881 & 3795822 & 5617703 & 9.201e-02 & 6.496e-05 & 1.917e-01 & 8.847e-05 & 2.837e-01 & 1.013e-04 & 32.43 & 0.02 \\
2.024e-05 & 1803783 & 3659412 & 5463195 & 9.110e-02 & 6.467e-05 & 1.848e-01 & 8.723e-05 & 2.759e-01 & 1.005e-04 & 33.02 & 0.02 \\
2.560e-05 & 1781329 & 3522731 & 5304060 & 8.997e-02 & 6.430e-05 & 1.779e-01 & 8.595e-05 & 2.679e-01 & 9.952e-05 & 33.58 & 0.02 \\
3.237e-05 & 1745893 & 3335557 & 5081450 & 8.818e-02 & 6.372e-05 & 1.685e-01 & 8.411e-05 & 2.566e-01 & 9.816e-05 & 34.36 & 0.02 \\
4.095e-05 & 1699183 & 3124685 & 4823868 & 8.582e-02 & 6.295e-05 & 1.578e-01 & 8.193e-05 & 2.436e-01 & 9.647e-05 & 35.22 & 0.02 \\
5.179e-05 & 1632461 & 2868884 & 4501345 & 8.245e-02 & 6.181e-05 & 1.449e-01 & 7.910e-05 & 2.273e-01 & 9.419e-05 & 36.27 & 0.02 \\
6.551e-05 & 1532429 & 2514577 & 4047006 & 7.740e-02 & 6.005e-05 & 1.270e-01 & 7.483e-05 & 2.044e-01 & 9.063e-05 & 37.87 & 0.02 \\
8.286e-05 & 1420606 & 2178577 & 3599183 & 7.175e-02 & 5.800e-05 & 1.100e-01 & 7.032e-05 & 1.818e-01 & 8.667e-05 & 39.47 & 0.03 \\
1.048e-04 & 1287384 & 1854042 & 3141426 & 6.502e-02 & 5.541e-05 & 9.364e-02 & 6.547e-05 & 1.587e-01 & 8.211e-05 & 40.98 & 0.03 \\
1.326e-04 & 1124192 & 1508971 & 2633163 & 5.678e-02 & 5.201e-05 & 7.621e-02 & 5.963e-05 & 1.330e-01 & 7.631e-05 & 42.69 & 0.03 \\
1.677e-04 & 962656 & 1204877 & 2167533 & 4.862e-02 & 4.833e-05 & 6.085e-02 & 5.372e-05 & 1.095e-01 & 7.017e-05 & 44.41 & 0.03 \\
2.121e-04 & 798088 & 929122 & 1727210 & 4.031e-02 & 4.420e-05 & 4.693e-02 & 4.753e-05 & 8.723e-02 & 6.341e-05 & 46.21 & 0.04 \\
2.683e-04 & 653396 & 706397 & 1359793 & 3.300e-02 & 4.015e-05 & 3.568e-02 & 4.168e-05 & 6.868e-02 & 5.684e-05 & 48.05 & 0.04 \\
3.393e-04 & 520208 & 522159 & 1042367 & 2.627e-02 & 3.595e-05 & 2.637e-02 & 3.601e-05 & 5.264e-02 & 5.019e-05 & 49.91 & 0.05 \\
4.292e-04 & 403220 & 376527 & 779747 & 2.036e-02 & 3.174e-05 & 1.902e-02 & 3.069e-05 & 3.938e-02 & 4.371e-05 & 51.71 & 0.06 \\
5.429e-04 & 305060 & 266055 & 571115 & 1.541e-02 & 2.768e-05 & 1.344e-02 & 2.588e-05 & 2.884e-02 & 3.761e-05 & 53.41 & 0.07 \\
6.866e-04 & 225668 & 184115 & 409783 & 1.140e-02 & 2.386e-05 & 9.299e-03 & 2.157e-05 & 2.070e-02 & 3.199e-05 & 55.07 & 0.08 \\
8.685e-04 & 163741 & 125591 & 289332 & 8.270e-03 & 2.035e-05 & 6.343e-03 & 1.784e-05 & 1.461e-02 & 2.697e-05 & 56.59 & 0.09 \\
1.099e-03 & 115538 & 84539 & 200077 & 5.835e-03 & 1.712e-05 & 4.270e-03 & 1.465e-05 & 1.010e-02 & 2.248e-05 & 57.75 & 0.11 \\
1.389e-03 & 79906 & 56421 & 136327 & 4.036e-03 & 1.425e-05 & 2.850e-03 & 1.198e-05 & 6.885e-03 & 1.858e-05 & 58.61 & 0.13 \\
1.758e-03 & 54156 & 37243 & 91399 & 2.735e-03 & 1.174e-05 & 1.881e-03 & 9.738e-06 & 4.616e-03 & 1.523e-05 & 59.25 & 0.16 \\
2.223e-03 & 36204 & 24493 & 60697 & 1.828e-03 & 9.601e-06 & 1.237e-03 & 7.899e-06 & 3.066e-03 & 1.242e-05 & 59.65 & 0.20 \\
2.812e-03 & 24106 & 15933 & 40039 & 1.217e-03 & 7.837e-06 & 8.047e-04 & 6.372e-06 & 2.022e-03 & 1.010e-05 & 60.21 & 0.24 \\
3.556e-03 & 15704 & 10381 & 26085 & 7.931e-04 & 6.327e-06 & 5.243e-04 & 5.144e-06 & 1.317e-03 & 8.152e-06 & 60.20 & 0.30 \\
4.498e-03 & 10242 & 6709 & 16951 & 5.173e-04 & 5.110e-06 & 3.388e-04 & 4.136e-06 & 8.561e-04 & 6.573e-06 & 60.42 & 0.38 \\
5.690e-03 & 6726 & 4332 & 11058 & 3.397e-04 & 4.141e-06 & 2.188e-04 & 3.324e-06 & 5.585e-04 & 5.309e-06 & 60.82 & 0.46 \\
7.197e-03 & 4413 & 2756 & 7169 & 2.229e-04 & 3.355e-06 & 1.392e-04 & 2.651e-06 & 3.621e-04 & 4.275e-06 & 61.56 & 0.57 \\
9.103e-03 & 2817 & 1775 & 4592 & 1.423e-04 & 2.680e-06 & 8.965e-05 & 2.128e-06 & 2.319e-04 & 3.422e-06 & 61.35 & 0.72 \\
1.151e-02 & 1761 & 1158 & 2919 & 8.894e-05 & 2.119e-06 & 5.848e-05 & 1.719e-06 & 1.474e-04 & 2.728e-06 & 60.33 & 0.91 \\
1.456e-02 & 1149 & 735 & 1884 & 5.803e-05 & 1.712e-06 & 3.712e-05 & 1.369e-06 & 9.515e-05 & 2.192e-06 & 60.99 & 1.12 \\
1.842e-02 & 746 & 466 & 1212 & 3.768e-05 & 1.379e-06 & 2.354e-05 & 1.090e-06 & 6.121e-05 & 1.758e-06 & 61.55 & 1.40 \\
2.330e-02 & 463 & 303 & 766 & 2.338e-05 & 1.087e-06 & 1.530e-05 & 8.791e-07 & 3.869e-05 & 1.398e-06 & 60.44 & 1.77 \\
2.947e-02 & 293 & 191 & 484 & 1.480e-05 & 8.645e-07 & 9.646e-06 & 6.980e-07 & 2.444e-05 & 1.111e-06 & 60.54 & 2.22 \\
3.728e-02 & 186 & 128 & 314 & 9.394e-06 & 6.888e-07 & 6.465e-06 & 5.714e-07 & 1.586e-05 & 8.949e-07 & 59.24 & 2.77 \\
4.715e-02 & 119 & 83 & 202 & 6.010e-06 & 5.509e-07 & 4.192e-06 & 4.601e-07 & 1.020e-05 & 7.178e-07 & 58.89 & 3.46 \\
5.964e-02 & 83 & 52 & 135 & 4.192e-06 & 4.601e-07 & 2.626e-06 & 3.642e-07 & 6.818e-06 & 5.868e-07 & 61.48 & 4.19 \\
7.543e-02 & 52 & 35 & 87 & 2.626e-06 & 3.642e-07 & 1.768e-06 & 2.988e-07 & 4.394e-06 & 4.711e-07 & 59.77 & 5.26 \\
9.541e-02 & 30 & 19 & 49 & 1.515e-06 & 2.766e-07 & 9.596e-07 & 2.201e-07 & 2.475e-06 & 3.535e-07 & 61.22 & 6.96 \\
1.207e-01 & 20 & 14 & 34 & 1.010e-06 & 2.259e-07 & 7.071e-07 & 1.890e-07 & 1.717e-06 & 2.945e-07 & 58.82 & 8.44 \\
1.526e-01 & 14 & 7 & 21 & 7.071e-07 & 1.890e-07 & 3.535e-07 & 1.336e-07 & 1.061e-06 & 2.314e-07 & 66.67 & 10.29 \\
1.931e-01 & 12 & 3 & 15 & 6.061e-07 & 1.750e-07 & 1.515e-07 & 8.748e-08 & 7.576e-07 & 1.956e-07 & 80.00 & 10.33 \\
2.442e-01 & 6 & 2 & 8 & 3.030e-07 & 1.237e-07 & 1.010e-07 & 7.142e-08 & 4.040e-07 & 1.428e-07 & 75.00 & 15.31 \\
3.089e-01 & 3 & 2 & 5 & 1.515e-07 & 8.748e-08 & 1.010e-07 & 7.142e-08 & 2.525e-07 & 1.129e-07 & 60.00 & 21.91 \\
3.907e-01 & 3 & 0 & 3 & 1.515e-07 & 8.748e-08 &  &  & 1.515e-07 & 8.748e-08 & 100.00 &  \\
4.942e-01 & 1 & 0 & 1 & 5.051e-08 & 5.051e-08 &  &  & 5.051e-08 & 5.051e-08 & 100.00 &  \\
6.251e-01 & 1 & 0 & 1 & 5.051e-08 & 5.051e-08 &  &  & 5.051e-08 & 5.051e-08 & 100.00 &  \\
7.906e-01 & 1 & 0 & 1 & 5.051e-08 & 5.051e-08 &  &  & 5.051e-08 & 5.051e-08 & 100.00 &  \\
1.000e+00 & 0 & 0 & 0 &  &  &  &  &  &  &  &  \\
\end{tabular}
\end{table}
\end{landscape}

\newpage
\begin{landscape}
\begin{table}[htbp]
\centering
\tiny
\caption*{Supplementary Table 6. Dataset C (ETHUSDT 2023-12-27 to 2024-08-09) results part 2.}
\label{ST6}
\begin{tabular}{CCCCCCCCCCCC}
$\delta$ & pMean & diff & pGeom & geoChi2p & geoKSp & lamHat & expKSp & lamCiLow & lamCiHigh & pPred \\
\midrule
1.000e-05 & 0.32 & -0.32 & 0.32 & \textless{}0.0001 & \textless{}0.0001 & 0.03 & \textless{}0.0001 & 0.03 & 0.03 & 0.03 \\
1.265e-05 & 0.32 & -0.31 & 0.32 & \textless{}0.0001 & \textless{}0.0001 & 0.04 & \textless{}0.0001 & 0.04 & 0.04 & 0.04 \\
1.600e-05 & 0.32 & -0.31 & 0.32 & \textless{}0.0001 & \textless{}0.0001 & 0.05 & \textless{}0.0001 & 0.05 & 0.05 & 0.05 \\
2.024e-05 & 0.33 & -0.30 & 0.33 & \textless{}0.0001 & \textless{}0.0001 & 0.06 & \textless{}0.0001 & 0.06 & 0.06 & 0.06 \\
2.560e-05 & 0.34 & -0.30 & 0.34 & \textless{}0.0001 & \textless{}0.0001 & 0.07 & \textless{}0.0001 & 0.07 & 0.07 & 0.07 \\
3.237e-05 & 0.34 & -0.29 & 0.34 & \textless{}0.0001 & \textless{}0.0001 & 0.09 & \textless{}0.0001 & 0.09 & 0.09 & 0.09 \\
4.095e-05 & 0.35 & -0.28 & 0.35 & \textless{}0.0001 & \textless{}0.0001 & 0.11 & \textless{}0.0001 & 0.11 & 0.11 & 0.11 \\
5.179e-05 & 0.36 & -0.27 & 0.36 & \textless{}0.0001 & \textless{}0.0001 & 0.14 & \textless{}0.0001 & 0.14 & 0.14 & 0.13 \\
6.551e-05 & 0.38 & -0.25 & 0.38 & \textless{}0.0001 & \textless{}0.0001 & 0.17 & \textless{}0.0001 & 0.17 & 0.17 & 0.16 \\
8.286e-05 & 0.39 & -0.24 & 0.39 & \textless{}0.0001 & \textless{}0.0001 & 0.21 & \textless{}0.0001 & 0.21 & 0.21 & 0.19 \\
1.048e-04 & 0.41 & -0.22 & 0.41 & \textless{}0.0001 & \textless{}0.0001 & 0.25 & \textless{}0.0001 & 0.25 & 0.25 & 0.22 \\
1.326e-04 & 0.43 & -0.21 & 0.43 & \textless{}0.0001 & \textless{}0.0001 & 0.30 & \textless{}0.0001 & 0.29 & 0.30 & 0.26 \\
1.677e-04 & 0.44 & -0.19 & 0.44 & \textless{}0.0001 & \textless{}0.0001 & 0.34 & \textless{}0.0001 & 0.34 & 0.34 & 0.29 \\
2.121e-04 & 0.46 & -0.17 & 0.46 & \textless{}0.0001 & \textless{}0.0001 & 0.39 & \textless{}0.0001 & 0.39 & 0.40 & 0.33 \\
2.683e-04 & 0.48 & -0.15 & 0.48 & \textless{}0.0001 & \textless{}0.0001 & 0.45 & \textless{}0.0001 & 0.45 & 0.45 & 0.36 \\
3.393e-04 & 0.50 & -0.13 & 0.50 & \textless{}0.0001 & \textless{}0.0001 & 0.50 & \textless{}0.0001 & 0.50 & 0.51 & 0.40 \\
4.292e-04 & 0.52 & -0.12 & 0.52 & \textless{}0.0001 & \textless{}0.0001 & 0.56 & \textless{}0.0001 & 0.56 & 0.56 & 0.43 \\
5.429e-04 & 0.53 & -0.10 & 0.53 & \textless{}0.0001 & \textless{}0.0001 & 0.61 & \textless{}0.0001 & 0.61 & 0.61 & 0.46 \\
6.866e-04 & 0.55 & -0.08 & 0.55 & \textless{}0.0001 & \textless{}0.0001 & 0.66 & \textless{}0.0001 & 0.66 & 0.67 & 0.48 \\
8.685e-04 & 0.57 & -0.07 & 0.57 & \textless{}0.0001 & \textless{}0.0001 & 0.71 & \textless{}0.0001 & 0.71 & 0.72 & 0.51 \\
1.099e-03 & 0.58 & -0.05 & 0.58 & \textless{}0.0001 & \textless{}0.0001 & 0.76 & \textless{}0.0001 & 0.75 & 0.76 & 0.53 \\
1.389e-03 & 0.59 & -0.05 & 0.59 & \textless{}0.0001 & \textless{}0.0001 & 0.79 & \textless{}0.0001 & 0.79 & 0.80 & 0.55 \\
1.758e-03 & 0.59 & -0.04 & 0.59 & \textless{}0.0001 & \textless{}0.0001 & 0.82 & \textless{}0.01 & 0.81 & 0.83 & 0.56 \\
2.223e-03 & 0.60 & -0.04 & 0.60 & \textless{}0.0001 & \textless{}0.0001 & 0.84 & \textless{}0.01 & 0.83 & 0.85 & 0.57 \\
2.812e-03 & 0.60 & -0.03 & 0.60 & \textless{}0.0001 & \textless{}0.0001 & 0.86 & 0.17 & 0.85 & 0.87 & 0.58 \\
3.556e-03 & 0.60 & -0.03 & 0.60 & \textless{}0.0001 & \textless{}0.0001 & 0.87 & 0.58 & 0.86 & 0.88 & 0.58 \\
4.498e-03 & 0.60 & -0.03 & 0.60 & \textless{}0.0001 & \textless{}0.0001 & 0.88 & 0.61 & 0.86 & 0.90 & 0.58 \\
5.690e-03 & 0.61 & -0.02 & 0.61 & \textless{}0.0001 & \textless{}0.0001 & 0.90 & 0.45 & 0.88 & 0.92 & 0.59 \\
7.197e-03 & 0.62 & -0.02 & 0.62 & \textless{}0.0001 & \textless{}0.0001 & 0.92 & 0.55 & 0.90 & 0.95 & 0.60 \\
9.103e-03 & 0.61 & -0.02 & 0.61 & \textless{}0.0001 & \textless{}0.0001 & 0.92 & 0.89 & 0.89 & 0.96 & 0.60 \\
1.151e-02 & 0.60 & -0.03 & 0.60 & \textless{}0.0001 & \textless{}0.0001 & 0.90 & 0.90 & 0.86 & 0.95 & 0.59 \\
1.456e-02 & 0.61 & -0.02 & 0.61 & \textless{}0.0001 & \textless{}0.0001 & 0.92 & 0.97 & 0.87 & 0.97 & 0.61 \\
1.842e-02 & 0.62 & -0.02 & 0.62 & \textless{}0.0001 & \textless{}0.0001 & 0.94 & \textless{}0.0001 & 0.87 & 1.00 & 0.61 \\
2.330e-02 & 0.60 & -0.03 & 0.60 & \textless{}0.0001 & \textless{}0.0001 & 0.90 & \textless{}0.0001 & 0.82 & 0.98 & 0.59 \\
2.947e-02 & 0.61 & -0.03 & 0.61 & \textless{}0.0001 & \textless{}0.0001 & 0.89 & 0.65 & 0.79 & 0.99 & 0.59 \\
3.728e-02 & 0.59 & -0.04 & 0.59 & \textless{}0.0001 & \textless{}0.0001 & 0.89 & 0.61 & 0.76 & 1.02 & 0.59 \\
4.715e-02 & 0.59 & -0.04 & 0.59 & \textless{}0.0001 & \textless{}0.0001 & 0.89 & 0.96 & 0.73 & 1.05 & 0.59 \\
5.964e-02 & 0.61 & -0.02 & 0.61 & \textless{}0.0001 & \textless{}0.0001 & 0.95 & 0.64 & 0.75 & 1.16 & 0.61 \\
7.543e-02 & 0.60 & -0.03 & 0.60 & \textless{}0.0001 & \textless{}0.0001 & 0.95 & 0.83 & 0.69 & 1.21 & 0.61 \\
9.541e-02 & 0.61 & -0.02 & 0.61 & \textless{}0.0001 & \textless{}0.0001 & 0.89 & 0.69 & 0.57 & 1.21 & 0.59 \\
1.207e-01 & 0.59 & -0.04 & 0.59 & \textless{}0.0001 & \textless{}0.0001 & 0.94 & 0.95 & 0.53 & 1.35 & 0.61 \\
1.526e-01 & 0.67 & 0.03 & 0.67 & \textless{}0.0001 & \textless{}0.0001 & 1.02 & 0.91 & 0.49 & 1.56 & 0.64 \\
1.931e-01 & 0.80 & 0.17 & 0.80 & \textless{}0.0001 & \textless{}0.0001 & 1.31 & 0.99 & 0.54 & 2.08 & 0.73 \\
2.442e-01 & 0.75 & 0.12 & 0.75 & \textless{}0.0001 & \textless{}0.01 & 1.20 & 0.45 & 0.24 & 2.16 & 0.70 \\
3.089e-01 & 0.60 & -0.03 & 0.75 & \textless{}0.0001 & 0.07 &  &  &  &  &  \\
3.907e-01 & 1.00 & 0.37 & 1.00 &  &  &  &  &  &  &  \\
4.942e-01 & 1.00 & 0.37 & 1.00 &  &  &  &  &  &  &  \\
6.251e-01 & 1.00 & 0.37 & 1.00 &  &  &  &  &  &  &  \\
7.906e-01 & 1.00 & 0.37 & 1.00 &  &  &  &  &  &  &  \\
1.000e+00 &  &  &  &  & &  &  &  &  &  \\

\end{tabular}
\end{table}
\end{landscape}

\end{document}